\documentclass[sigconf, nonacm]{acmart}

\newcommand\vldbdoi{XX.XX/XXX.XX}
\newcommand\vldbpages{XXX-XXX}
\newcommand\vldbvolume{14}
\newcommand\vldbissue{1}
\newcommand\vldbyear{2020}
\newcommand\vldbauthors{\authors}
\newcommand\vldbtitle{\shorttitle} 
\newcommand\vldbavailabilityurl{https://github.com/DataManagementLab/eleet}
\newcommand\vldbpagestyle{plain}

\newcommand*\circled[1]{\textcircled{\raisebox{-0.9pt}{#1}}}

\newcommand{\method}{\texttt{ELEET}}
\newcommand{\idattr}{\emph{document-level key}}
\newcommand{\singlerow}{\emph{single-row}}
\newcommand{\Singlerow}{Single-row}
\newcommand{\Multirow}{Multi-row}

\usepackage{algorithm}
\usepackage{algpseudocode}
\usepackage{cleveref}
\usepackage{relsize}
\usepackage{amsmath}
\usepackage{caption}
\usepackage{subcaption}
\usepackage{array}
\usepackage{multirow}
\let\tablemultirow\multirow
\renewcommand{\multirow}{\emph{multi-row}}

\DeclareMathOperator{\concat}{\mathbin\Vert}

\begin{document}
\title{\method{}: Efficient Learned Query Execution over Text and Tables}
\subtitle{[Technical Report --- Extended Version of \cite{eleet}]}

\author{Matthias Urban}
\affiliation{
  \institution{Technical University of Darmstadt}
}
\email{matthias.urban@cs.tu-darmstadt.de}
\author{Carsten Binnig}
\affiliation{
  \institution{Technical University of Darmstadt \& DFKI}
}
\email{carsten.binnig@cs.tu-darmstadt.de}

\begin{abstract}
{In this paper, we present \method{}, a novel execution engine that allows one to seamlessly query and process text as a first-class citizen along with tables.}
To enable such a seamless integration of text and tables, \method{} leverages learned multi-modal operators (MMOps) such as joins and unions that seamlessly combine structured with unstructured textual data.
While large language models (LLM) such as GPT-4 are interesting candidates to enable such learned multi-modal operations,  we deliberately do not follow this trend to enable MMOps, since it would result in high overhead at query runtime.
Instead, to enable MMOps, \method{} comes with a more efficient small language model (SLM) that is targeted to extract structured data from text.
Thanks to our novel architecture and pre-training procedure, the \method{}-model enables high-accuracy extraction with low overheads.
{In our evaluation, we compare query execution based on \method{} to baselines leveraging LLMs such as GPT-4 and show that \method{} can speed up multi-modal queries over tables and text by up to $575\times$ without sacrificing accuracy.}
\end{abstract}

\maketitle

\pagestyle{\vldbpagestyle}
\begingroup\small\noindent\raggedright\textbf{PVLDB Reference Format:}\\
\vldbauthors. \vldbtitle. PVLDB, \vldbvolume(\vldbissue): \vldbpages, \vldbyear.\\
\href{https://doi.org/\vldbdoi}{doi:\vldbdoi}
\endgroup
\begingroup
\renewcommand\thefootnote{}\footnote{\noindent
This work is licensed under the Creative Commons BY-NC-ND 4.0 International License. Visit \url{https://creativecommons.org/licenses/by-nc-nd/4.0/} to view a copy of this license. For any use beyond those covered by this license, obtain permission by emailing \href{mailto:info@vldb.org}{info@vldb.org}. Copyright is held by the owner/author(s). Publication rights licensed to the VLDB Endowment. \\
\raggedright Proceedings of the VLDB Endowment, Vol. \vldbvolume, No. \vldbissue\ %
ISSN 2150-8097. \\
\href{https://doi.org/\vldbdoi}{doi:\vldbdoi} \\
}\addtocounter{footnote}{-1}\endgroup

\ifdefempty{\vldbavailabilityurl}{}{
\vspace{.3cm}
\begingroup\small\noindent\raggedright\textbf{PVLDB Artifact Availability:}\\
The source code, data, and/or other artifacts have been made available at \url{\vldbavailabilityurl}.
\endgroup
}


\section{Introduction}

\noindent\textbf{More than Tables.} Decades of research have turned relational databases into highly optimized systems for managing tabular data.
However, modern data applications need to deal with other data modalities as well that are often used in addition to tabular data, such as texts or image data \cite{symphony,neuraldb,wannadb}.
Unfortunately, traditional relational databases are not well-equipped to handle these multi-modal scenarios.
Instead, practitioners are forced to process modalities other than tables outside the database or integrate them by manually transforming such modalities into tabular form first.

\noindent\textbf{Query Execution over Multi-modal Data.} 
At the same time, rapid advancements in natural language processing and computer vision have made it easier to extract insights from texts, images, as well as other modalities.
In light of these developments, we believe it is time to bring these innovations to the world of databases and enable users to seamlessly query multi-modal data.
{
Although some extensions have been integrated into commercial database systems such as full-text search or pattern matching for textual data \cite{full-text-search-microsoft}, modalities such as text do by far not allow for the same level of querying as tabular data.
}
Our work aims to fill this gap.
Hence, we propose \method{}, an approach that leverages query plans with learned operators that allow us to seamlessly process data of modalities other than tables as if they were available in tabular form.

\begin{figure}[t]
 \centering
 \includegraphics[width=0.85\linewidth]{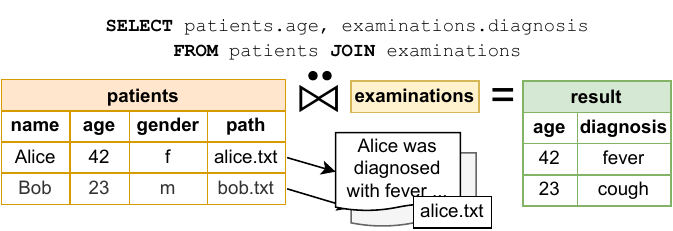}
 \caption{Example of a query that executes a multi-modal join between a patient table and examination reports.
 \method{} analyzes the texts and extracts values for each queried attribute, such as the diagnosis from each examination report.}
 \label{fig:teaser}
\end{figure}

\noindent\emph{A Simple Example.}
Figure \ref{fig:teaser} illustrates how we envision how \method{} can be used by applications.
In the example, the database stores structured patient information (using a table) alongside textual patient reports that contain additional diagnostic information per patient.
If the diagnostic information were stored in tabular form as well, a SQL query, as shown at the top of Figure \ref{fig:teaser}, could easily be used to analyze the correlation between the patient's age and her diagnosis.
{
If the information is, however, stored inside textual reports, today a data scientist would need to write many lines of code to create a data extraction pipeline that retrieves the diagnostic information from text.}
Only after extraction would it then be possible to query this information at a similar level to tabular data.

\begin{figure*}
  \centering
  \includegraphics[width=0.87\linewidth]{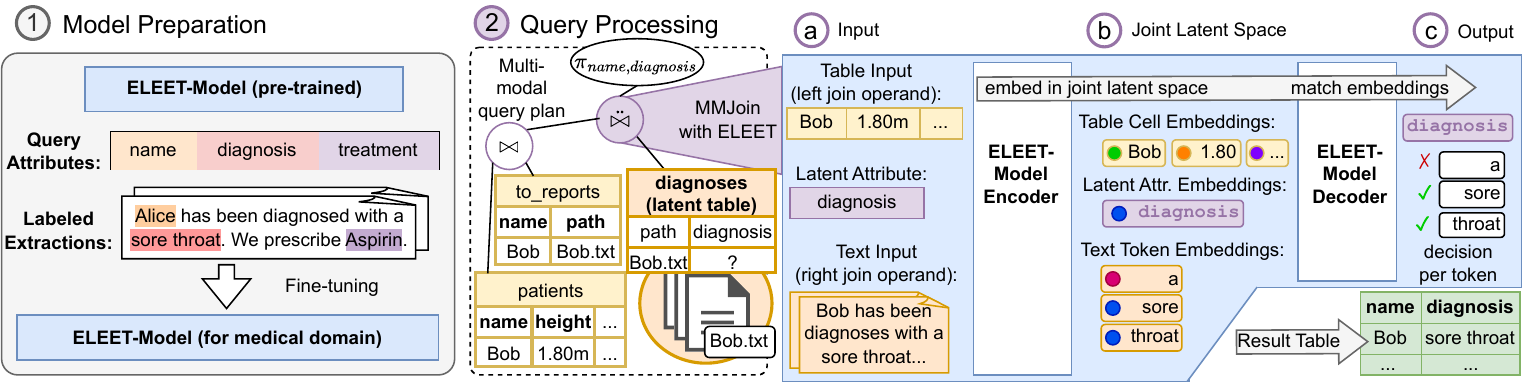}
    \caption{Overview of \method{}. In an offline phase, the \method{}-model can be fine-tuned for unseen domains \circled{1}.
    Fine-tuning the \method{}-model for an unseen domain is a one-time effort and requires a small sample of a few labeled texts.
    \circled{2} For query execution, \method{} uses multi-modal query plans that contain traditional (white) and multi-modal database operators (purple).
    To compute the result of a multi-modal operation such as a join over texts, the \method{}-model is used (see \circled{a} to \circled{c}): During the execution of a multi-modal operation, the \method{} model first computes embeddings of the query attributes, texts, and table input \circled{a}, using its encoder \circled{b}.
    Afterwards, the \method{}-model matches text token embeddings to query attribute embeddings to extract the output table from the text using its extractive decoder \circled{c}, which decides which tokens qualify for a given query attribute.}
  \label{fig:overview}
\end{figure*}

\noindent\textbf{Learned Multi-Modal Operators.}
{
The goal of our work is to challenge this need for a ``special treatment'' for modalities other than tables, allowing users to query them seamlessly and declaratively as if they were tables.
}
To enable seamless querying of multi-modal data, we propose to extend relational query plans with so-called learned multi-modal operators (MMOps).
The basic idea of MMOps is that they extend the set of operators used in traditional query engines by new operators that can natively process data sources of other modalities.
As shown in Figure \ref{fig:teaser}, for example, a multi-modal join operator for tables and texts allows users of \method{} to join the patient table directly with the linked patient reports.
As such, the data analyst can correlate the relationship between the patient's age and their diagnoses in a simple and efficient manner.

\noindent\textbf{Multi-Modal Query Plans.}
The rationale behind MMOps, such as the multi-modal join, is that they can accept data sources of other modalities as input and produce tables as an output.
Thus, MMOps nicely integrate with the existing query processing capabilities of traditional databases since MMOps can be composed into query plans along with relational operators to enable complex analytical queries.
For example, after the multi-modal join operator shown in Figure \ref{fig:teaser}, other relational operators, such as a projection or a filter, can be applied to provide rich query functionality to users.
Moreover, as we elaborate later in the paper, in addition to multi-modal joins, \method{} implements a wide spectrum of different MMOps, including multi-modal scans, unions, and aggregations.

\noindent\textbf{Realizing Multi-Modal Operators using LLMs?}
The key idea in realizing such multi-modal operators for text and tables is to build on recent advances in the area of language models.
In fact, recent language models (e.g., GPT-3 \cite{gpt3}, GPT-4 \cite{openaiGPT4TechnicalReport2023a}, LLaMA \cite{touvronLLaMAOpenEfficient2023a, touvronLlamaOpenFoundation}, PaLM \cite{chowdheryPaLMScalingLanguage2022a, anilPaLMTechnicalReport2023} or Gemini \cite{gemini}) have shown remarkable results on a wide range of text-processing tasks.
While it has been shown that recent language models can be used out-of-the-box to transform a text into a table, we argue that such language models are not readily usable for efficient query execution.
In particular, due to their immense size (i.e., GPT-4 has more than a trillion parameters), they are very computationally expensive.
Each call for a single text can take several seconds, leading to query runtimes of multiple minutes even for small text collections as we show in our experiments.

\noindent\textbf{Small Language Models to the Rescue?}
Therefore, in this paper, we take a different route: instead of building \method{} on large language models (LLMs) such as GPT, we instead base multi-modal operators on a small language model (SLM) to achieve a more efficient execution of multi-modal operators.
A key to this SLM is that we use a model architecture that targets table extraction from text and pre-train the model to learn the essential skills to perform MMOps.
Thanks to this pre-training, the \method{}-model can provide high accuracy and efficiency at the same time.
For example, our model architecture avoids using costly autoregressive decoding, which is prominent in LLMs and requires many passes through the model to construct the output rows from texts.
Instead, our model uses an extractive approach using embeddings of text and query attributes, which can extract data from text in a single model pass.

\noindent\textbf{More than table extraction from text.}
Finally, a last important property of our model is that it can incorporate additional signals from tabular data sources when extracting structured data from text, which can improve the extraction quality.
For example, when executing a multi-modal join between a table and a text collection, as discussed before in our example, the multi-modal join can take structured data (e.g., the name of a patient) as input to extract the correct diagnosis from the text.
This might be particularly helpful in scenarios where the text contains information about multiple patients.
Moreover, providing such signals from structured data can also reduce the runtime of multi-modal joins, as we discuss later.

\noindent\textbf{Contributions.}  To summarize, in this paper, we present three major contributions:
(1) As the core contribution, we present the \method{}-model. For realizing the \method{}-model we use a novel SLM that targets table extraction from text.
For pre-training the SLM, we construct a new parallel pre-training corpus of tables and texts.
(2) As a second contribution, we show how MMOps can be realized using a pre-trained \method{}-model.
We present a wide range of different MMOps such as a multi-modal scan that can turn text collections into tables or more complex operations like joins, unions, selections, and aggregations operating on texts and tables.
(3) Finally, we provide an extensive evaluation using four data sets to challenge our approach and evaluate how accurate and efficient multi-modal query plans are executed with \method{}.
The evaluation uses data sets from different domains and a wide range of multi-modal query plans.
Using these workloads, we compare \method{} against strong baselines, including some that leverage LLMs such as GPT-3 or 4.

\noindent\textbf{Outline.}
We first provide an overview of \method{} in Section \ref{sec:overview}, {before
we define its data model and algebra in Section \ref{sec:definitions}}.
Then, we explain the details of the \method{}-model in Section \ref{sec:model}
and how it can be used to realize MMOps in Section \ref{sec:mmops}.
Finally, we present our evaluation in Section \ref{sec:eval},
related work in Section \ref{sec:related-work} and conclude in Section \ref{sec:conclusion}.

\section{Overview of \method{}} \label{sec:overview}

In the following, we first explain the overall procedure of executing queries with \method{}.
After that, we discuss its key design principles.

\subsection{Overall Procedure}

\method{} executes a multi-modal query plan containing traditional database operators and MMOps.
Such a multi-modal query plan consumes data from tables and text collections that can be seamlessly treated as tables (called latent tables) using our \method{}-model.
\method{} supports different MMOps in multi-modal query plans.
For instance, the query  $\pi_{\mathrm{name}, \mathrm{diagnosis}}((\mathrm{patients} \Join \mathrm{to\_reports}) \ddot \Join \mathrm{diagnoses})$ in Figure \ref{fig:overview} uses a multi-modal join to combine text with table data.
In the following, we sketch how this join can be realized by using our \method{}-model in Figure \ref{fig:overview} \circled{2}.

\noindent\textbf{A Sketch of a Multi-Modal Join.}
The join in Figure \ref{fig:overview} needs to extract the diagnosis for each patient tuple coming from the first join of the \emph{patients} and \emph{to\_reports} table (i.e., each patient can have multiple reports).
For executing this query, we feed the attributes to extract from text (i.e., \emph{diagnosis}; called \emph{latent attribute}) together with the patient data from the first join
and the text documents to be joined into the \method{}-model.
For example, for joining the patient tuple of Bob with his patient report in Figure \ref{fig:overview} \circled{a}, \method{} feeds the patient tuple (containing name, height, ...), the latent attribute \emph{diagnosis}, and the patient report of Bob into the \method{}-model.
For extracting the diagnosis, the encoder of our \method{}-model maps all inputs into a joint latent space \circled{b}.
Afterwards, the decoder identifies spans of texts in the report that qualify as diagnosis, such as the text span \emph{sore throat} in Figure \ref{fig:overview} \circled{c}.
Finally, the result row $\{\mathrm{name} \mapsto \mathrm{Bob}, \mathrm{diagnosis} \mapsto \mathrm{sore\;throat}\}$ with the extracted values from the text is materialized.
One assumption we make in \method{} is that there exists a foreign key relationship between the tuples in the relational table and the text collection, i.e., patient reports are linked to a patient tuple.
{In Section \ref{sec:definitions} we discuss the data model of \method{} more formally}.

\noindent\textbf{Fine-Tuning for unseen Domains.}
While the \method{} model is pre-trained to learn table extraction from text, it clearly benefits from fine-tuning on texts of domains unseen during pre-training (see \circled{1} in Figure \ref{fig:overview}).
To do so, in a model preparation phase (offline), the user labels a few example texts by marking text spans that are extractions for potential query attributes.
Based on a pre-trained \method{}-model, only a few fine-tuning samples are necessary.
In our evaluation, we show that only a small number of labeled documents are typically sufficient to achieve high accuracy for unseen domains.

\subsection{Key Design Principles}\label{sec:design_principles}

\textbf{Efficiency of Extraction.} The efficiency of query execution is a major concern for \method{}.
We tackle this using three key design principles for the \method{}-model:
(1) Firstly, regarding the \method{}-model, we use a small language model with only $140$ million parameters that we optimize for performing query operations.
The \method{}-model is multiple orders of magnitudes smaller than recent LLMs (e.g., GPT-3 has $175$ billion and GPT-4 has $1.76$ trillion parameters) and thus provides much lower inference latencies.
{
(2) Secondly, in our model architecture as we discuss next, we avoid costly autoregressive decoding that LLMs typically use, which requires many passes through a transformer-based decoder and thus leads to high inference times as shown in Section \ref{sec:ablation-model}.
}
Instead, the \method{}-model uses an extractive approach that computes its output in a single pass through the model.
(3) Finally, beyond the model itself, optimal physical operator implementations of MMOps in \method{} can help to improve the efficiency of query execution further.
For example, for a multi-modal selection that applies a filter on attributes from the texts, we leverage an index-based implementation to avoid the high scan cost of scanning the full text for all documents in the collection.

{\noindent\textbf{Accuracy without Regrets.}
Another key design principle of \method{} is that we do not trade efficiency for accuracy.
Instead, as we specialize our model for the task of extracting structured data from text, it is highly accurate on this task and often even more accurate than much larger general-purpose language models such as GPT-4.
We achieve this through our pre-training procedure, which teaches the model the necessary skills to perform table extraction from text. This allows \method{} to be highly accurate while being more efficient than LLMs, as we show in our evaluation.
}

\noindent\textbf{Online and Offline Execution}
While this paper aims to enable online execution of multi-modal query plans, \method{} can also be used to pre-compute extractions from text offline (i.e., constructing a materialized view of a multi-modal query).
However, we think there are many scenarios where the ability to execute multi-modal query plans online is crucial.
For example, in a setting where text collections are continuously updated, online query execution is important to provide up-to-date query results and can avoid the high cost of view maintenance.
Moreover, online query execution can also be attractive in a setting where queries only need to process a few texts, and materializing a table of a potentially huge text collection would cause high overheads.
In these cases, online processing saves the additional high storage and pre-processing overheads of materialization.
Finally, materialization prevents ad-hoc queries where query attributes are not known in advance.

{
\section{Data Model and Algebra Definition} \label{sec:definitions}
In this section, we explain which types of queries are supported by \method{} by formally defining its data model and \method{}'s algebra.

\subsection{Data Model of \method{}} \label{sec:data-model}
The data model of \method{} builds upon the relational data model and extends it by text collections and latent tables.

\begin{table}[h!]
\centering
\small
\caption{Overview of \method{}'s data model.}
\begin{tabular}{c | c}
 \toprule
 Symbol & Name \\ [0.5ex]
 \midrule
 $T_A$ & Table with attributes $A$ \\
 $D$ & Text collection \\
 $D.LT_{LA}$ & Latent table with latent attributes $LA$ \\
 \bottomrule
\end{tabular}
\label{tab:data_model}
\end{table}

\noindent\textbf{Text collections.}
A text collection $D = \{d_1, \dots, d_{|D|}\}$ is a set of text documents.
Similar to how different rows in database tables follow the same schema defining a set of attributes $A$  per table,
we assume that different documents $d$ of a text collection $D$ also expose the same attributes.
For instance, in a text collection of patient reports, each document contains information about the patient's name and the diagnosis.
Furthermore, each document is uniquely identified by its file path $path(d)$ and can contain an arbitrary number of tokens: $d=w_1 \dots w_{|d|}$.
In \method{}, text collections can be queried standalone but can also be linked to traditional tables.
Traditional tables can be linked to text collections by storing their file paths as an additional attribute in the tables, essentially creating a foreign key relationship between tables and text collections.
In fact, many real-world datasets today already use this model to link text documents and tables.
For example, we analyzed the GitTables corpus \cite{gittables}, a collection of 1M tables, and found that ~15k tables already contain paths to externally stored text files. 

\noindent\textbf{Latent Tables.}
In \method{}, users register latent tables over a text collection $D$ to make it available for query processing.
Similar to regular tables, the user has to define a schema $LA = \{la_1,\dots,la_{|LA|}\}$ for a latent table $D.LT$. 
We assume that the user knows the data well and can specify a reasonable schema.
We denote such a latent table as $D.LT_{LA}$.
The schema defines the attributes that can be extracted from text (e.g., name and diagnosis of medical reports).
A tuple $t \in D.LT$ thus contains a value $v = w^v_1 \dots w^v_{|v|}$ that can be extracted from text $d = w_1 \dots w^v_1 \dots w^v_{|v|} \dots w_n $ for each attribute in $LA$.
Furthermore, each tuple contains the file path to the document $d$.
Importantly, defining a latent table does not yet extract any values from the text.
Instead, a latent table is merely a handle that can be used in query plans.
The values are extracted on the fly during query execution as explained in Section \ref{sec:design_principles}.

\noindent\textbf{\Singlerow{} and \Multirow{} Latent Tables.}
Finally, in \method{} we distinguish between \singlerow{} and \multirow{} latent tables.
A \singlerow{} latent table is where the user knows that each document $d\in D$ contributes exactly one tuple $t$ to a latent table $D.LT$ (e.g., each patient report always contains a single diagnosis and treatment).
On the other hand, a \multirow{} latent table is the general case where each document can contribute an arbitrary number of tuples to a latent table.
In case of a \multirow{} latent table, the user has to define a \idattr{} $la_{key} \in LA$ such that each latent tuple $t$ coming from the same document $d$ is uniquely defined by the values for $la_{key}$.
For example, if a medical report contains information about multiple diagnoses, the diagnosis name would be a sensible \idattr{}.
Note that defining the schema, the type of latent table (\singlerow{} vs. \multirow{}) and the \idattr{} needs to be done only once per latent table and not per document, thus causing the same overhead as creating a schema for a normal table.

{
Moreover, we found that some of the decisions mentioned above, like choosing the type of latent table or choosing the \idattr{} can be automated using LLMs.
For example, to decide whether a latent table is \multirow{}, we prompt an LLM with a few example texts together with the schema of the latent table and ask it whether each text contributes one or multiple rows to the latent table.
Since choosing the type of latent table is done only once per latent table, as explained before, using a costly LLM is reasonable.
See our experiments in Section \ref{sec:auto-registration} for details on the prompts and our findings.

}

\subsection{Algebra of \method{}}

\method{} uses an algebra to compose multi-modal query plans.
The algebra of \method{} extends the traditional relational algebra with multi-modal operators summarized in Table \ref{tab:algebra} and formally defined in Section \ref{sec:formal-definitions}.

\noindent\textbf{Multi-modal Scan.}
The most important operator is the multi-modal scan operator.
A multi-modal scan takes a latent table $D.LT_{LA}$ as input and materializes a normal table $T_{LA \cup \{\mathrm{path}\}}$ as output by extracting values $v = w^v_1 \dots w^v_{|v|}$ for each latent attribute from all text documents.
The output table of a multi-modal scan can thus be used as input to normal relational operators such as joins and filters.
Important is that the multi-modal scan is sufficient for expressing all possible multi-modal queries in \method{}.
As such, the other multi-modal operators in Table \ref{tab:algebra} do not enrich the expressivity of queries in \method{} but instead improve the quality of query results or the efficiency of query execution (or both).
In the following, we briefly explain each multi-modal operator's interface, its role, and how it can optimize a multi-modal query plan.
More details about the individual operators can be found in Section \ref{sec:mmops}.

\begin{table}
    \centering
    \small
    \caption{Summary of the multi-modal operators.}
    \begin{tabular}{c | c | c}
        \toprule
        Name & Expression & Output Type \\ [0.5ex]
        \midrule
        Scan & $ \ddot Scan(D.LT_{LA})$ & $T_{LA \cup \{\mathrm{path}\}}$ \\ 
        Join & $T_A \; \ddot\Join \; D.LT_{LA}$ & $T_{A \cup LA}$ \\
        Union & $T_A \; \ddot\cup \; D.LT_{LA}$ & $T_{A}$ \\
        Projection & $\ddot\pi_{LA' \subseteq LA}(D.LT)$ & $D.LT_{LA'}$ \\
        \tablemultirow{2}{*}{Selection} & $\ddot\sigma_{cond}(D.LT_{LA})$ & $D.LT_{LA}$ \\
        & $\ddot\sigma_{cond}(T_A)$ & $T_A$ \\
        Aggregation & $\ddot\chi_{F, A' \subseteq A}(T_A)$ & $T_A$ \\
        \bottomrule
    \end{tabular}
    \label{tab:algebra}
\end{table}

\noindent\textbf{Multi-modal Join.} The second operator is a multi-modal join $T \ddot \Join D.LT$, which can replace a combination of a scan with a traditional join $T \Join_{T.path = D.LT.path} \ddot Scan(D.LT)$.
For the multi-modal join, the table $T$ must be linked to the document collection $D$ by storing file $path$ as an additional attribute in $T$, which is used as a join key.
The task of the multi-modal join is to extract values from text linked to a tuple in $T$ by leveraging attribute values from the tuple.
The multi-modal join  $\mathrm{patients} \, \ddot \Join \, \mathrm{reports.examinations}$, for example, extracts for each patient (stored in a table) values for the latent attributes of the latent \emph{examinations} table (e.g., diagnosis, treatment, etc.)  from the textual reports.

The multi-modal join is an optimization over using 
$\mathrm{patients} \, \Join \, \ddot Scan(\mathrm{reports.examinations})$ 
as the multi-modal join can use the data in the patients table during extraction from the textual reports.
For example, the patient table containing the patient name can help extract the relevant parts of the medical report text, which is particularly helpful if the reports contain information about multiple patients.

\noindent\textbf{Multi-modal Union.} The third multi-modal operation is the union $T \ddot \cup D.LT$, which can be used to replace a traditional union and a scan $T \cup \ddot Scan(D.LT)$.
For instance, when a hospital stores its patient information in tabular form, while another stores it as reports, one could combine both with a multi-modal union.
Important is that the attributes exposed by the normal and latent table are compatible in types, meaning they store the same type of information (e.g., both store patient name and diagnosis).
Again, the difference to a scan is that the union can use the tabular data $T$ as additional context for the model.
In the case of a union, this additional context is example values (e.g., example names and diagnoses) for each latent attribute, which can help extract values from the text.

\noindent\textbf{Multi-modal Projection \& Selection.}  
The multi-modal projection $\ddot \pi_{LA'}(D.LT)$
projects the columns from a latent table without materializing it.
As such, it is an optimization over using traditional projection after a multi-modal scan $\pi_{LA'}(\ddot Scan(D.LT))$, which would need to materialize values for all columns.
In contrast, the multi-modal selection $\ddot\sigma_{cond}(D.LT)$ as shown in Table \ref{tab:algebra} (first variant) reduces the number of texts using a filter condition $cond$ on a latent attribute.
Important is that the text documents are filtered without materializing them as output table. As such, it is an optimization over using traditional selection after a multi-modal scan $\sigma_{cond}(\ddot Scan(D.LT))$ which would first need to extract rows for all documents before filtering.
Moreover, another feature of the multi-modal selection is that it improves selection quality since it can detect matches of text values similar to the value used by the filter condition $cond$ (e.g., diagnosis=fever can also match the synonym high temperature in text).
The second variant of the multi-modal selection $\ddot\sigma_{cond} \ddot Scan(T)$ is different since it uses a normal table as input. However, it can still detect matches of similar values in filter predicates and table values. This selection can be used in case a selection can not be pushed down to the text. An example is a disjunctive selection on the output of a multi-modal join, which filters based on attributes of the table and the latent table (e.g., patient.age=18 OR reports.examinations.diagnosis='fever'). 

\noindent\textbf{Multi-modal Aggregation.}  
Finally, similar to the second selection, multi-modal aggregation $\ddot \chi_{F,A'}(T)$ operates on a normal table but can group similar values for attributes coming from text extractions.
It replaces a traditional aggregation over a multi-modal scan or join, which is less robust if extracted values contain synonyms.
Like traditional aggregation, multi-modal aggregation takes parameters $F$ for the aggregation function over one of the attributes in $T$ and $A'$ for the group-by attributes.

}

\subsection{Algebraic Rewrites in \method{}}
\label{appendix:b1}

\noindent As discussed before, interfaces of multi-modal operators are designed so they can often be used to optimize query execution of multi-modal query plans both in terms of runtime and accuracy.
More specifically, we imagine a query compiler that, given a user query (e.g., in SQL), will first instantiate a query plan that only contains multi-modal scans and traditional operators.
Since the scan output is a normal table, and afterwards traditional operators can be composed arbitrarily, the system supports arbitrarily complex SQL queries where data sources can be text exposed as latent tables and normal tables.
However, in an optimization step, the query compiler can look for patterns in the query plan where the plan can be optimized by inserting another multi-modal operation (e.g., multi-modal joins, unions, filters, etc.).
For instance, if the user inputs the SQL query from Figure \ref{fig:teaser}, the query compiler would first instantiate the multi-modal query $\pi_{\mathrm{age}, \mathrm{diagnosis}} (\mathrm{patients} \Join \mathrm{Scan}(\mathrm{examinations}))$, that contains a multi-modal scan on the latent table \emph{examinations}.
Afterwards, the compiler would notice that using the data in the \emph{patients} table helps extract the diagnoses of the correct patients from the text.
Moreover, materializing all columns of the \emph{examinations} table using the scan is also unnecessary because the user is only interested in diagnoses.
Hence, the compiler replaces the join and the scan with a multi-modal join that extracts values from text while leveraging the context from the tabular operand.
Afterwards, it pushes down the projection and replaces it with a multi-modal projection, preventing all columns from being materialized.
Hence, the result query plan that is optimized with multi-modal operators is: $\pi_{\mathrm{age},\mathrm{diagnosis}}(\mathrm{patients} \, \ddot \Join \, \ddot \pi_{diagnosis}(examinations))$.
Similarly, multimodal unions could be inserted when users try to union a table with a latent table, and the quality of extractions can be improved with example values; multi-modal selections could be inserted if the user filters based on values extracted from text, and so on, as explained above.
This example illustrates that \method{} supports arbitrary queries, and multi-modal operators can be used in many cases for optimization.
However, while this shows the potential of using  multi-modal operators to enhance the accuracy and efficiency of multi-modal plans, 
building an optimizer for multi-modal plans that uses cost models for estimating performance (and accuracy) is beyond the scope of this paper but represents an interesting opportunity for future research on multi-modal databases.

{
\subsection{Formal Definition of Multi-Modal Operators} \label{sec:formal-definitions}

So far, we have restricted the formal definitions of the operators to their input and output sets (i.e., the domain and co-domain of the underlying function). In this section, we formally define the expected result of a multi-modal operator given its input data.

\noindent\textbf{Multi-modal Scan (\Singlerow{} Latent Table).} For the scan of a \singlerow{} latent table, the scan extracts for all latent attributes $la \in LA$, the "correct" values (i.e., as given by the ground truth) from each text $d \in D$. The output table $R$ contains one output tuple per input text document $d$:
$$
R = \left\{
\begin{array}{l}
 \{\mathrm{path} \mapsto path(d)\}      \\
  \cup \; \{la \mapsto \mathrm{extract}(d, la)  | la \in LA \}
\end{array}
\middle\vert \; d \in D \right\}
$$
In this definition, a single output tuple $t \in R$ is extracted from each document $d \in D$ by extracting a value $v = extract(d, la)$ for all latent attributes $la \in LA$.
The expression extract($d, la$) denotes the extraction of a value $v$ for the latent attribute $la \in LA$ (e.g., diagnosis) from document $d$ and is a sequence of tokens
$$v = w^v_1 \dots w^v_{|v|}$$ (e.g. $v =$ sore throat, $w_1^v$ = sore, $w_2^v$ = throat).
The values are extracted directly from text $d$, meaning the token sequence of $v$ appears in $d$ as explained before:
$$d = w_1 \dots w^v_1 \dots w^v_{|v|} \dots w_n$$

Moreover, all result tuples $t\in R$ contain the special \emph{path} attribute that contains the file path $path(d)$ of document $d$.
The file path of a document $d$ indicates where $d$ is stored in the file system and uniquely identifies each document.
This allows for further operations on the scan output, such as selections based on the file path.
Another important property enabled by the file path is that we can join the output with a relevant table.
As explained in Section \ref{sec:data-model}, a table $T$ can contain links to a text collection $D$ by explicitly storing the file paths of text documents in one of its columns.
For example, a patient table could store the file path to each patient's patient report.
That way, a traditional table $T$ (e.g., the patient table) can be joined with the output of a scan on a text collection (e.g., patient reports),

by joining on the path column using a traditional join operator: $T \Join_{T.path = D.LT.path} \ddot Scan(D)$.
However, \method{} also offers the multi-modal join as a more accurate and efficient alternative, which we define later.

\noindent\textbf{Multi-modal Scan (\Multirow{} Latent Table).} The definition is different for a \multirow{} latent table as each document can contribute multiple tuples to the output table.
To enable this, the user has to define a \idattr{} $la_{key}$ as explained in Section \ref{sec:definitions}.

The \idattr{} is a special latent attribute that allows us to define an output table that consists of arbitrarily many tuples per input text.
The \idattr{} can have multiple values for a single input text, and each of those values represents a different output tuple.
For example, in a document collection of patient reports, where the same report can contain the diagnosis of multiple patients, $la_{key} = \mathrm{patient\_name}$ is a sensible \idattr{}.
In this example, if three patients with patient\_name Alice, Bob, and Carol were mentioned in a single text, this would result in three output tuples for that text according to the definition below:

In this case, extract$(d, \mathrm{patient\_name}) = \{\mathrm{Alice}, \mathrm{Bob}, \mathrm{Carol}\}$ is a set of all patients mentioned in the one document $d$.
Moreover, since each patient has their own diagnosis, we also need to define extract$(d, v_{key}, la)$, which is the value for latent attribute $la$ given the value $v_{key}$ for the \idattr{}.
For instance, if the diagnosis of Bob is fever, then extract$(d, \mathrm{Bob}, \mathrm{diagnosis}) = \mathrm{fever}$.
As such, we can define the expected scan output $R$ as:
$$
R = \left\{ 
\begin{array}{l}
\{\mathrm{path} \mapsto path(d)\} \\
\cup \, \{la_{key} \mapsto v_{key}\} \\
\cup \, \{la \mapsto \mathrm{extract}(d, v_{key}, la) \\ \;\;\;\; | \; la \in LA   \setminus \{la_{key}\}\}
\end{array}
\middle\vert \, 
\begin{array}{l}
     d \in D  \; \; \land \\
     v_{key} \in extract(d, la_{key})
\end{array}
\right\}    
$$
Like before, as we can see in the expression left of the vertical bar, each tuple $t \in R$ contains the special \emph{path} attribute that contains the file path $path(d)$ of the document $d$.
Additionally, it also contains the \idattr{} $la_{key}$ as one of its attributes (e.g., $la_{key}$ = patient\_name) and one of its potential values $v_{key}$ (e.g., Alice or Bob or Carol from the example before) is assigned to it.
Finally, the output tuple contains values for all other latent attributes $la \in LA \setminus \{la_{key}\}$  extracted from text $d$.
However, different from before, we can see in the predicate right of the vertical bar that for a single document $d$, the expected output contains multiple output tuples, one for each $v_{key}$ mentioned in the document (e.g., one tuple for each patient if multiple patients are contained in the report).

\noindent\textbf{Multi-Modal Join.} As described in Section \ref{sec:definitions}, the multi-modal operators beyond the multi-modal scan do not make multi-modal queries more expressive.
Instead, they are optimizations to improve accuracy of the extractions.
For example, a multi-modal join improves the extraction accuracy by using the tabular context of the structured table in the input.

To be more formal, the expected output of a multi-modal join is defined to be equivalent to the combination of a scan and a traditional join: $$T \ddot\Join D.LT=T \Join_{T.path=D.LT.path} \ddot Scan(D.LT)$$
As such, the formal definition of the multi-modal join is based on the formal definition of the multi-modal scan and \emph{it is not necessary to provide a separate formal definition}. 

\noindent\textbf{Multi-Modal Union.} Similarly, the expected output of a multi-modal union is equivalent to a traditional union with the output of a scan:
$$T \ddot\cup D.LT=T \cup \ddot Scan(D.LT)$$
Again, the formal definition of the multi-modal union is based on the formal definition of the multi-modal scan and thus \emph{it is not necessary to  provide a separate formal definition}. 

\noindent\textbf{Multi-Modal Selection and Aggregation.}
As explained in Sections \ref{sec:selection} and \ref{sec:aggregation}, the multi-modal selection and the multi-modal aggregation operator operate on normal tables as input (and not text documents).
These operators aim to make the selection and aggregation more robust than their traditional counterparts by considering semantic synonyms.
For instance, if a user selects all tuples with diagnosis \emph{fever}, we also want to return patients with diagnosis \emph{high body temperature}.
Similarly, if the user aggregates with the group-by key \emph{diagnosis}, we want to place \emph{fever} and \emph{high body temperature} in the same group.
As such, \emph{the definition of the multi-modal aggregation and the multi-modal selection is equivalent to their traditional counterparts} and again does not require an additional formal definition.
}
\section{The \method{}-model}
\label{sec:model}

At the core of \method{} is the \method{}-model which is used for extracting structured tuples from text.
In the following, we show the model architecture and how the model is pre-trained.

\begin{figure}[t]
  \centering
  \includegraphics[width=\columnwidth]{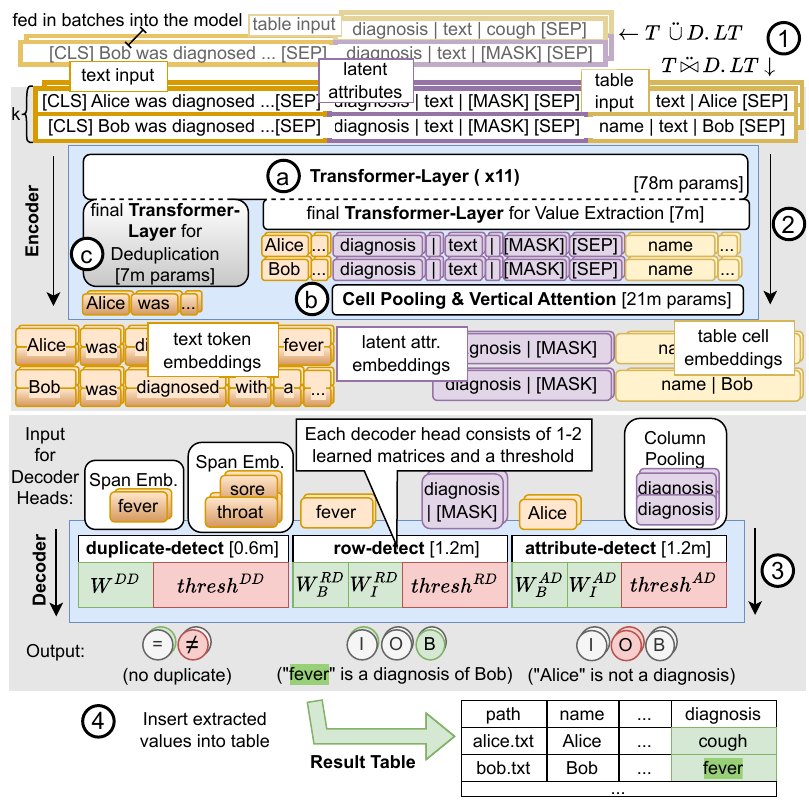}
       \caption{Model architecture. After encoding a batch of sequences that each contain an input text, the latent attributes, and traditional table values \circled{1} using 12 (11+1) transformer layers \circled{2} \circled{a},
    all embeddings corresponding to the same cells or latent attributes are pooled, before vertical attention lets signal flow between groups of k rows \circled{b}.
    A separate final transformer layer computes a second set of text embeddings optimized for detecting duplicates \circled{c}.
    The decoder \circled{3} consists of several heads for the different sub-tasks for extracting table data from texts.
    For instance, the row-detect head is used to find extractions in the text. For this, it pairs the embedding of each text token with the embedding of a masked cell (i.e., the attribute to be extracted) and classifies whether the token is part of the attribute or not.
    The tokens that are marked to be extracted are inserted into the output table \circled{4}.
}
    
  \label{fig:model}
\end{figure}

{
\subsection{Model Architecture}

Our \method{}-model uses an encoder-decoder architecture as depicted in Figure \ref{fig:model} that follows the design principles outlined in Section \ref{sec:design_principles}.
In our model, the encoder first computes embeddings from the input.
The decoder then consists of several lightweight decoder heads that use these embeddings to perform the different subtasks for transforming texts to tables (e.g., value extraction, deduplication).
For the sake of presentation, we explain the design of our model and how it can be used to execute multi-modal joins or unions on a \singlerow{} latent table.
For \singlerow{} latent tables, for each text document, we only need to extract a single value for each latent attribute (i.e., for each patient report, we need to extract one diagnosis, one treatment, etc.).
More details about all operations and \multirow{} latent tables are explained in Section \ref{sec:mmops}.
}

\noindent\textbf{Encoder.}
The encoder computes embeddings for the inputs required to extract values from text documents. 
{
Figure \ref{fig:model} \circled{1} and Algorithm \ref{alg:encoder} shows an example of a multi-modal join.
The input to the encoder consists of the text tokens coming from a text document (orange) and the latent attributes $LA$ for which values need to be extracted from text (purple), which is the diagnosis attribute in Figure \ref{fig:model}.
Additional non-latent context can be fed into the model (yellow) that comes from the tabular operand of a multi-modal join or union (this is generally optional).
For the multi-modal join example, the additional context is the tuples linked to the textual report.
In Figure \ref{fig:model}, we see two tuples (e.g., Alice and Bob) and two separate reports shown as text input (i.e., one per patient) from which we aim to extract the diagnoses.
For unions, we use two rows randomly picked from the tabular operand of the union.
The \method{}-model can use the additional values from the table as context for extractions, as we explain below.
We use the same linearization of \texttt{name | type | value} for the latent attributes and non-latent table values but use a MASK token for latent attributes to indicate that a value needs to be extracted for them, as shown in Figure \ref{fig:model} \circled{1}.
}
As shown in Figure \ref{fig:model} \circled{2}, after linearization, the inputs are fed into the 12 transformer layers (see \circled{a})
to compute their representations (i.e., embeddings).
{
Multiple input sequences are fed into the model in a batch to increase efficiency.
}
After running the input through 11 transformer layers, there are different paths for different decoder subtasks.
In particular, the paths use different final transformer layers:

For \emph{extracting values}, after running the individual rows through the transformer layers and obtaining embeddings for all input tokens, we pool all token embeddings belonging to the same table cell
(i.e., the embeddings for attribute name, type, and value / MASK) into one cell embedding (cell pooling).
{
For unions, it is important that signal flows from the rows containing example values to the embeddings of the masked cells to compute optimal embeddings for extraction.
Hence, we apply vertical self-attention \cite{tabert} across different rows, i.e., across the cell embeddings of the same column  (see \circled{b} in Figure \ref{fig:model}).
In \method{}, we feed the sequences in groups of $k=3$ into vertical self-attention \cite{tabert}.
Hence, for unions, we pair all sequences with MASK tokens and text with both sequences containing example values, meaning we extract values for a single text per vertical self-attention call. 

}

The second path is for \emph{detecting duplicates}, which is needed if text mentions the same extraction multiple times (e.g., fever and high body temperature for the same patient).
For this, we use a final transformer layer that is separate from the one for extracting values (see \circled{c} in Figure \ref{fig:model}).
This is because the two decoder subtasks benefit from embeddings of different characteristics.
For value extraction, embeddings of text-tokens should be similar to those of the latent attributes for which they are a value.
For detecting semantic duplicates, however, embeddings of text tokens that belong to different semantic concepts should be dissimilar, even though they are a value for the same latent attribute.
Hence, this separate final transformer layer produces embeddings that the decoder can use to detect duplicate values more easily.
{
To summarize, the encoder computes two embeddings per text token $w$ (one for value extraction $\hat w^{VE}$ and one for duplicate detection $\hat w^{DD}$) and one embedding $\hat c_{i, j}$ per cell, where $1 \leq i \leq k$ and $1 \leq j \leq |LA|$.
{
\begin{figure}[t]
\begin{algorithm}[H]
\caption{
    \circled{1} Encoder of \method{} for a multi-modal join. Scans and union differ in the table data that is fed in the model as explained in Section \ref{sec:model}.
    }\label{alg:encoder}
\begin{algorithmic}[1]
    
    \Require texts $d_1 \dots d_k$, latent attributes $LA$, table tuples $t_1 \dots t_k$ (table tuples come from the tabular operand of the join; $t_i$ is linked to $d_i$ via the file path)
    \State Feed the tuple-text pairs into 11 Transformer layers \circled{a}
        \begin{equation*}
        \begin{split}
        \hat{E} \gets \mathrm{Transformer[\times11]}(d_1 \concat & [SEP] \concat \mathrm{linearized}(LA) \concat \mathrm{linearized}(t_1), \\
        \cdots \\
        d_k \concat & [SEP] \concat \mathrm{linearized}(LA) \concat \mathrm{linearized}(t_k))
        \end{split}
        \end{equation*}
    \State $\hat{E}^{DD} \gets \mathrm{Transformer}[\times1]_{DD}(\hat E)$  \Comment Embeddings for DD \circled{c}
    \State \Comment Discard non-text token embeddings
    \State $(\hat w^{DD}_{1,1}, \dots, w^{DD}_{k, n}), \_ \gets \hat E^{DD}$  
    \State $\hat{E} \gets \mathrm{Transformer}[\times1]_{RD, AD}(\hat E)$  \Comment Embeddings for RD, AD
    \State \Comment Separate in text and table embeddings
    \State $(\hat w^{VE}_{1,1}, \dots, w^{VE}_{k, n}), (\hat e_{1,1}, \dots \hat e_{k,\bar m}) \gets \hat E$  
    \State \Comment Take the mean of all embeddings belonging to the same cell
    \State $(\hat c_{1,1}, \dots \hat c_{k,m}) \gets Cell-Pool((\hat e_{1,1}, \dots \hat e_{k,\bar m}))$ \Comment{\circled{b}}
    \State $(\hat c_{1,1}, \dots \hat c_{k,m}) \gets \mathrm{Vertical-Attention}((\hat c_{1,1}, \dots \hat c_{k,m}))$

    \State \Return $(\hat w^{VE}_{1,1}, \dots, w^{VE}_{k, n}), (\hat w^{DD}_{1,1}, \dots, w^{DD}_{k, n}), (\hat c_{1,1}, \dots \hat c_{k,m})$
\end{algorithmic}
\end{algorithm}
\end{figure}
}
}

\noindent\textbf{Decoder.}
The decoder (Figure \ref{fig:model} \circled{3}) generates the output table by extracting a value for each latent attribute per text.
It uses three lightweight decoder heads and the embeddings from the encoder.

{
For materializing the output of a join, for example, it is important to extract only values for the entities described by the tuple linked to each text.
For instance, when we join a patient tuple (e.g., Bob) with a textual report of this patient, we typically only want to extract the diagnosis of Bob from the text, even when other patients are mentioned in the text.
To do that, we use the \emph{row-detect (RD)} head that is pre-trained to extract only values for the particular entity mentioned in the input of the model (e.g., Bob) as given by the tuple from the table.
For extracting the output of a union, we want to use example values and use \emph{attribute-detect (AD)} as we discuss below.
{
\begin{figure}[t]
\begin{algorithm}[H]
\caption{
    \circled{2} Decoder of \method{}: \textbf{Row Detect}.
    The actual implementation computes I-O-B tags for all input sequences at once using matrix multiplication instead of a loop.
    {\textbf{Attribute Detect} works analogously, using $\hat {a_j} = \sum_{i=1}^{k} \hat c_{i, j}$ instead of attribute embedding $\hat c$ and weights $W_\mathrm{tag}^{AD}, thresh^{AD}$ instead of $W_\mathrm{tag}^{RD}, thresh^{RD}$.}
    }\label{alg:decoder:row_detect}
\begin{algorithmic}[1]
    
    \Require text $d = w_1 \dots w_n$, token embeddings $\hat w_1^{VE} \dots \hat w_n^{VE}$, masked cell embedding $ {\hat c}$
    \State $(\mathrm{tags}_1, \dots, \mathrm{tags}_n) \gets (O, \dots, O)$ \Comment Initialize tags
    \ForAll{tokens $w_i \in w_1 \dots w_n$ with embedding $\hat w_i$}
        \State $\mathrm{tags}_i \gets \begin{array}{c}
            \mathrm{argmax} \\
            \scriptstyle{\mathrm{tag} \in \{I, O, B\}}
        \end{array}
            \begin{cases}
                \hat w_i^T \cdot W_{\mathrm{tag}}^{RD} \cdot \hat c & if\; \mathrm{tag} \in \{B, I\}\\
                thresh^{RD} & else
            \end{cases}$
    \EndFor
    \State \Return I-O-B-decode($d$, tags) \Comment Return a tag for extraction
\end{algorithmic}
\end{algorithm}
\end{figure}
}

Both the RD and the AD head extract values from the text to fill in values for latent attributes (e.g., diagnosis) and thus use the embeddings for extracting values $\hat w^{VE}$ (\circled{b}).
The RD head pairs the embedding of each text token $\hat w^{VE}$ (orange in Figure \ref{fig:model}) with the embedding of a masked cell $\hat c$ of a latent attribute (purple) and classifies whether the token is part of a value for the cell.
It consists of matrices $W^{RD}_I$, $W^{RD}_B$ and a learned threshold $thresh^{RD}$ to perform the classification according to Algorithm \ref{alg:decoder:row_detect}.
{We employ so-called \emph{I-O-B (Inside-Outside-Beginning) classification} to extract potentially multiple tokens for each attribute.
With I-O-B tags, the first text token for a value is marked with a B-tag, and subsequent tokens belonging to the same value receive an I-tag.
Otherwise, tokens are marked with an O-tag.}
{
The AD head works identically as the RD head but first pools all cell embeddings $\hat c_{i,j}$ of the same latent attribute into an attribute embedding $\hat {a_j} = \frac{1}{k} \sum_{i=1}^{k} \hat c_{i, j}$.
}
Then, it classifies based on this embedding, learned weights $W_I^{AD}, W_B^{AD}$ and learned threshold $thresh^{AD}$ which tokens are a value for that latent attribute, independent of the entity.

Finally, to avoid extracting the same value twice (e.g., to avoid duplicate rows in a \multirow{} latent table, as we explain later), it is necessary to check whether multiple extracted spans (i.e., all tokens extracted for a latent attribute) refer to the same concept.
For this, we compute span embeddings $\hat v$ according to \citet{span-representation} from the text token embeddings coming from the final transformer layer for deduplication $\hat w^{DD}$ (\circled{c})
and then classify which spans are duplicates using the \emph{duplicate-detect (DD)} head.
The DD head consists of a learned matrix $W^{DD}$ and a learned threshold $thresh^{DD}$ and considers two spans $v_A$ and $v_B$ as duplicates if the learned similarity $\hat v_A^T \cdot W^{DD} \cdot \hat v_B$ is larger than $thresh_{DD}$.
The procedure to find duplicates is shown in Algorithm \ref{alg:decoder:duplicate_detect}.

{
\begin{figure}[t]
\begin{algorithm}[H]
\caption{
    \circled{2} Decoder of \method{}: \textbf{Duplicate Detect}.
    }\label{alg:decoder:duplicate_detect}
\begin{algorithmic}[1]
    
    \Require Spans $V=\{v_A, v_B, \dots\}$, token embeds. $\hat w^{DD}_1 \dots \hat w^{DD}_n$
        \State $\hat v_A, \hat v_B, \dots \gets$ span\_embeddings\cite{span-representation}($v_A, v_B, \dots$; $\hat w^{DD}_1 \dots \hat w^{DD}_n$) 
        \State similarities = $[\hat v_A, \hat v_B, \dots]^T \cdot W^{DD}\cdot[\hat v_A, \hat v_B, \dots] - thresh^{DD}$
        \State \Comment We use agglomerative clustering with a distance threshold to find groups (clusters) of semantically equivalent values.
        \State $\bar V \gets$ Agglom.cluster(spans=$V$, dist=-similarites, dist\_thresh=0)
        \State \Comment Return longest span per cluster.
        \State \Return $\{\mathrm{argmax}_{v \in \mathrm{cluster}} |v| \; | \; \mathrm{cluster} \in \bar V\}$
\end{algorithmic}
\end{algorithm}
\end{figure}
}

After the values for all latent cells have been extracted and deduplicated, they can be inserted into the latent table to produce the materialized output.
However, in the general case of \multirow{} latent tables, where each text can contribute multiple tuples to the latent table, only value extraction is insufficient.
We explain the algorithm to support \multirow{} latent tables in Section \ref{sec:mmops}.
}

\subsection{Pre-training} \label{sec:pretraining}

{
The functionality of each of the three decoder heads represents a skill of the \method{} model that is necessary to perform MMOps as specified in the algebra in Section \ref{sec:definitions}.
}
These skills are independent of concrete data sets and thus should be taught to the model during pre-training.
For pre-training the \method{} model, we pair the encoder with our decoder heads and train them end-to-end.
{
However, we do not start pre-training from scratch but start with the pre-trained weights of TaBERT \cite{tabert} for the transformer and vertical self-attention layers.
These are the model components that also exist in TaBERT.
For the decoder layers, which do not have a counterpart in TaBERT, we start with randomly initialized weights.
During TaBERT's pre-training, the model sees pairs of texts and tables.
Hence, the resulting weights are a good starting point for \method{}.
We show this and the additional importance of our pre-training in our ablation study in Section \ref{sec:ablation}.}

\noindent\textbf{Skill 1: Align latent attributes to text (AD head).}
To support that our model learns to detect all values for certain attributes, we introduce the \emph{Attribute-Text-Alignment (ATA)} task, which aligns table attributes to text.
In this task, we pair texts with tables and use the AD head to detect segments of text that are a potential value for each attribute.
We use the labels of our pre-training corpus to compute a cross-entropy classification loss $\mathcal{L}_{ATA}$.

\noindent\textbf{Skill 2: Extract values for table rows (RD head).}
For joins, as explained before, it is important to extract only values belonging to a given table row (e.g., only extract values for a given patient).
To let our model learn which values correspond to which table row (e.g., in texts about multiple entities), we introduce the \emph{Masked Cell Reconstruction (MCR)} pre-training task.
In this task, we pair table rows and texts, mask out random cell values of table rows (that occur in both text and table row), and use the RD head to reconstruct the masked values from the text.
To do this, the RD head leverages signals from neighboring cells, including those from the same row.
MCR thus teaches the model to extract values for a certain row only (e.g., only the diagnoses of Bob), as required for the RD head.
For this pre-training objective, we use the labels of our pre-training corpus to compute a cross-entropy classification loss $\mathcal{L}_{MCR}$ for classifying the I-O-B tags in Algorithm \ref{alg:decoder:row_detect}.

\noindent\textbf{Skill 3: Detect duplicates (DD head).}
Finally, we introduce the \emph{Duplicate-Detection (DD)} pre-training task to let the model learn to detect whether two spans refer to the same concept.
For each attribute, \emph{DD} takes pairs of text spans as input and is trained to predict whether they are the same.
We train the DD head by classifying whether pairs of spans are duplicates using the labels of our pre-training corpus to compute a cross-entropy loss $\mathcal{L}_{DD}$.

\noindent\textbf{Combined Pre-Training.}
We use a combined pre-training to add up all the losses of all objectives and train the entire model (including the encoder) using this multi-task loss.
This has also shown benefits in other transformer-based models \cite{bert, strug} where multiple pre-train objectives are used.
To balance the losses, we use a weighted sum. We choose $\alpha=300$, $\beta=80$, $\gamma=\delta=1$ by examining the performance on a development set.

\begin{equation*}
    \mathcal{L} = \alpha\cdot \mathcal{L}_{MCR} + \beta\cdot\mathcal{L}_{ATA} + \gamma\cdot\mathcal{L}_{DD} + \delta\cdot\mathcal{L}_{MLM}
\end{equation*} \label{eq:loss}

Moreover, for the pre-training, we realized that our model benefits from adding the original Masked Language Model loss $\mathcal{L}_{MLM}$ of BERT \cite{bert}.
We thus added $\mathcal{L}_{MLM}$ to the combined loss.
Finally, to ensure that the model also utilizes signal from the table values during pre-training and not only the schema information (i.e., attribute names) of the table, we randomly mask out attribute names from the linearized input to our model in 15\% of the cases (which is a fraction we empirically validated to provide the best performance).
{
The complete procedure is shown in Algorithm \ref{alg:pretraining}.
}

{
\begin{figure}[t]
\begin{algorithm}[H]
\caption{
    Compute pre-training loss value for a single sample of k tuple-text pairs.
    In practice, multiple samples are fed into the model in a single batch, and the implementation avoids for-loops in favor of efficient matrix multiplications.
    }\label{alg:pretraining}
\begin{algorithmic}[1]
    
    \Require texts $d_1 \dots d_k$, table tuples $t_1 \dots t_k$, labels $Y^{RD}, Y^{AD}, Y^{DD}$, loss weights $\alpha, \beta, \gamma, \delta$, MLM-loss $\mathcal{L}_{MLM}$
    \State \textbf{Feed tuple-text pairs into Encoder to get all embeddings}
    \begin{equation*}
        \begin{array}{c}
             \hat{w}^{VE}_{1,1} \\
              \ddots \\
             \hat{w}^{VE}_{k, n} \\
        \end{array},
        \begin{array}{c}
             \hat{w}^{DD}_{1,1} \\
             \ddots \\
             \hat{w}^{DD}_{k, n} \\
        \end{array},
        \begin{array}{c}
               \hat{c}_{1,1} \\
              \ddots \\
               \hat{c}_{k, m} 
        \end{array}
        \gets \mathrm{Encoder} \left(
        \begin{array}{c}
            d_1 \concat [SEP] \concat \mathrm{linearized}(t_1), \\
            \cdots \\
            d_k \concat [SEP] \concat \mathrm{linearized}(t_k) \\
        \end{array}
        \right)
    \end{equation*}
    \State $\mathcal{L}_{MCR} \gets \mathcal{L}_{CTA} \gets \mathcal{L}_{DD} \gets 0$
    \ForAll{$i \in 1 \dots k$, masked cell $c \in t_i$, token $w \in d_i$}
    \State $l_I \gets \hat w^T \cdot W_I^{RD} \cdot \hat c$ \Comment Compute $\mathcal{L}_{MCR}$
    \State $l_B \gets \hat w^T \cdot W_B^{RD} \cdot \hat c$
    \State $l_O \gets thresh_{RD}$
    \State $\mathcal{L}_{MCR} \gets \mathcal{L}_{MCR} - \sum\limits_{x \in \{I, O, B\}}\mathbf{1}_{Y^{RD}_{i,w,c}=x} \cdot \log{\frac{\exp{l_{x}}}{\sum_{x' \in \{I, O, B\}} \exp l_{x'}}}$
    \EndFor
    \ForAll{$i \in 1 \dots k$, $j \in 1 \dots m$, token $w \in d_i$}
        \State $\hat a \gets \frac{1}{k} \sum_{i'=1}^k{\hat c_{i', j}}$ \Comment Compute column embedding
        \State $l_I\gets \hat w^T \cdot W_I^{AD} \cdot \hat a$ \Comment Compute $\mathcal{L}_{CTA} $
        \State $l_B \gets \hat w^T \cdot W_B^{AD} \cdot \hat a$
        \State $l_O \gets thresh_{AD}$
        \State $\mathcal{L}_{CTA} \gets \mathcal{L}_{CTA} - \sum\limits_{x \in \{I, O, B\}}\mathbf{1}_{Y^{AD}_{i, j, w}=x} \cdot \log{\frac{\exp{l_{x}}}
                                          {\sum_{x' \in \{I, O, B\}} \exp l_{x'}}}$
    \EndFor
    \ForAll{$i \in 1 \dots k$, values $v_{A}, v_{B} \in d_i$ (given by $Y^{AD}$)}
        \State $\hat v_A, \hat v_B, \dots \gets$ span\_embed.\cite{span-representation}($v_A, v_B, \dots$; $\hat w^{DD}_{i,1} \dots \hat w^{DD}_{i,n}$) 
        \State $l_Y\gets \hat v_A^T \cdot W^{DD} \cdot \hat v_B$  \Comment Compute $\mathcal{L}_{DD} $
        \State $l_N \gets thresh_{DD}$
        \State $\mathcal{L}_{DD} \gets \mathcal{L}_{DD} - \sum_{x \in \{Y, N\}}\mathbf{1}_{Y^{DD}_{v_A, v_B}=x} \cdot \log{\frac{\exp{l_{x}}}
                                                      {\sum_{x' \in \{Y, N\}} \exp l_{x'}}}$
    \EndFor
    \State $\mathcal{L}_{MCR} \gets \frac{1}{N_{MCR}} \cdot \mathcal{L}_{MCR}$  \Comment{Compute mean of loss values}
    \State $\mathcal{L}_{CTA} \gets \frac{1}{N_{CTA}} \cdot \mathcal{L}_{CTA}$
    \State $\mathcal{L}_{DD} \gets \frac{1}{N_{DD}} \cdot \mathcal{L}_{DD}$
    \State $\mathcal{L} \gets \alpha\cdot \mathcal{L}_{MCR} + \beta\cdot\mathcal{L}_{CTA} + \gamma\cdot\mathcal{L}_{DD} + \delta\cdot\mathcal{L}_{MLM}$
    \State \Return $\mathcal{L}$  \Comment{Train using AdamW \cite{adamw} optimizer}
\end{algorithmic}
\end{algorithm}
\end{figure}
}

{
\noindent\textbf{A New Pre-Training Corpus}
To pre-train our model using the above procedure, we created a new open-domain pre-training corpus with Wikipedia abstracts as texts, tables constructed from Wikidata, and labels from T-REx \cite{trex}.
We describe the corpus in Section \ref{sec:corpus}.

}

\begin{figure}[t]
\begin{algorithm}[H]
\caption{\Multirow{} multi-modal scan for a single document.}\label{alg:complex}
\begin{algorithmic}[1]
    \Require $d=(w_1, \dots, w_n)$, $LA=(la_{key}, la_1, \dots, la_m)$ \\
    \textbf{1. Get values} for $la^{key}$
    \State $\hat{w}_1 \dots \hat{w}_n, \hat{c}_{la_{key}}, \hat{c}_1, \dots, \hat{c}_m \gets \mathrm{Encoder}(d \concat [SEP] \concat \mathrm{lin.}(LA))$
    \State $V^{key} \gets \mathrm{attribute-detect}((\hat{w}_1, \dots, \hat{w}_n), \hat{c}_{la_{key}})$
    \State $\{v_1^{key}, \dots, v_l^{key}\} \gets \mathrm{duplicate-detect}(V^{key})$ \\
    \textbf{2. Get values for $la_1, \dots, la_m$}
    \State $\hat W, \hat C \gets \mathrm{Encoder}(d \concat [SEP] \concat \mathrm{lin.}(v_1^{key}) \concat \mathrm{lin.}(la_1, \dots, la_m) \; , \dots$
    \State \hspace{7.2em}               $d \concat [SEP] \concat \mathrm{lin.}(v_l^{key}) \concat \mathrm{lin.}(la_1, \dots, la_m))$
    \For{$i=1$ to $l$}
    \State $V_{i,1}, \dots, V_{i,m} \gets \mathrm{row-detect}((\hat{w}_{i,1}, \dots, \hat{w}_{i,n}), \hat{c}_{i,1}) \; , \dots$
    \State \hspace{6.4em}                 $\mathrm{row-detect}((\hat{w}_{i,1}, \dots, \hat{w}_{i,n}), \hat{c}_{i,m})$
    \State $\{v_{i,1}\}, \dots, \{v_{i,m}\} \gets \mathrm{duplicate-detect}(V_{i,1})\;, \dots$
    \State \hspace{8.1em}                        $\mathrm{duplicate-detect}(V_{i,m})$
    \EndFor
    \State \Return $\begin{bmatrix}
        v_1^{key} & v_{1,1} & \dots & v_{1,m}\\
        \vdots & \vdots & \ddots & \vdots \\
        v_l^{key} & v_{l,1} & \dots & v_{l, m}
    \end{bmatrix}$

\end{algorithmic}
\end{algorithm}
\end{figure}

{
\section{Multi-Modal Operations}
\label{sec:mmops}

In the previous section, we introduced the \method{}-model by demonstrating how it can be used for multi-modal joins for \singlerow{} latent tables.
In this section, we discuss how the details of all operators of the \method{} algebra can be realized using the \method{}-model.}

\begin{figure}[t]
  \centering
  \includegraphics[width=0.9\linewidth]{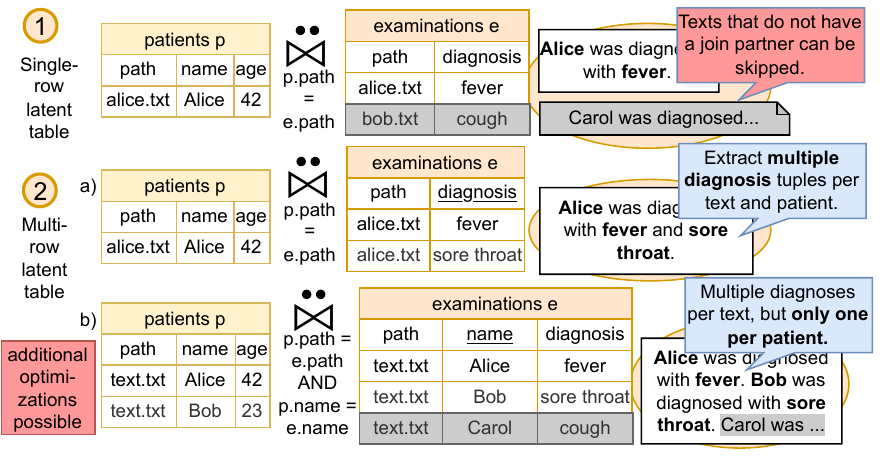}
  \caption{
      Besides the join with a \singlerow{} latent table \circled{1}, there are two cases for \multirow{} latent tables \circled{2}.
      In the first case, multiple tuples need to be extracted per table row of the tabular operand (a).
      An interesting special case is when for each table tuple, only a single tuple needs to be extracted (b).
      Joins allow for several automatic optimizations, depending on the case.
      For example, texts not having a join partner can be skipped during extraction, which is particularly beneficial when the table has been filtered before the join.
      Moreover, in the case of (b), there is no need to run Algorithm \ref{alg:complex}.
  }
  \label{fig:join-cases}
\end{figure}

\subsection{Multi-Modal Scans} \label{sec:scan}

The scan is the most important operator that turns a latent table into a normal one by extracting the values for all latent attributes from each text.
For implementing the scan of a \singlerow{} latent table, we feed in each document together with the latent attributes into our encoder: 
the input sequences have the form $d \concat [SEP] \concat \mathrm{linearized}(LA)$, where we use MASK tokens for linearizing the latent attributes as before.
{

Hence, unlike the join and union explained in Section \ref{sec:model}, we cannot access any tabular context.
Therefore, we set $k=1$ for scans, disabling vertical signal flow between input rows via vertical self-attention and mean pooling across rows to obtain attribute embeddings.
Hence, $\hat a_j$ is simply set to the masked cell embedding $\hat c_{1, j}$.
After feeding the sequences in the encoder, we use the attribute-detect (AD) head to extract a value for every latent attribute from each text using Algorithm \ref{alg:decoder:row_detect} with matrices $W_I^{AD}, W_B^{AD}$ and threshold $thresh^{AD}$ for I-O-B tagging.
}

Next, we explain the scan on a \multirow{} latent table.
In a \multirow{} latent table,  a single text $d\in D$ may correspond to multiple tuples $t_{d,1}, t_{d,2}, \dots$.
For instance, a patient report can document multiple diagnoses of a patient, resulting in multiple output tuples of a scan.
As described before, the user needs to define a \idattr{} $la_{key}$ for the \multirow{} latent table, which is required to distinguish between the different tuples coming from the same text.
For the patient reports in the example before, we assume the attribute $la^{key}=$\texttt{diagnosis} serves as the \idattr{}, meaning that all rows coming from the same text should represent a different diagnosis.
For realizing a multi-modal scan on such a table, we use an Algorithm \ref{alg:complex} which is composed of two phases.

In the first phase of the procedure, the model extracts all values for the \idattr{} $la_{key}$, using our AD head in the decoder (to extract all diagnoses; see lines 2+3 of Algorithm \ref{alg:complex}).
The number of values extracted in this step determines the number of output rows for a given text $d$.
Moreover, we use the duplicate-detect head to avoid generating duplicate rows when the text mentions values for $la_{key}$ twice (e.g., when the same text mentions both \emph{fever} and \emph{high body temperature}; see line 4 of Algorithm \ref{alg:complex}).

In the second phase, the extraction process is conducted on the remaining attributes that depend on the \idattr{}, denoted as $la_1, la_2, \dots$. 
To accomplish this, the MASK token of the \idattr{} $la_{key}$ is now replaced with the values extracted in the first phase (line 6).
For the remaining attributes (which still have a MASK in the input), we extract the values using the \method{} model.
Here, we use the row-detect (RD) head to extract only values corresponding to those extracted in the first phase (line 9).

To process texts longer than our model's context window of 512 tokens, we use a sliding window to process the whole text.
The first phase is executed independently using the sliding windows to extract all keys.
Afterwards, we collect and deduplicate the extracted values for the \idattr{} across windows.
Finally, the second phase can process each window independently with the extracted keys.
Afterwards, all values are collected across windows to generate the output tuples.
{ 
In an experiment in Section \ref{sec:breakdown-textlength}, we show that this approach works well even for longer texts.
}

{\subsection{Multi-Modal Joins and Unions}\label{sec:join_union}
{
As introduced in Section \ref{sec:definitions}, multimodal joins ($T \,\ddot\Join\, D.LT$) and unions ($T\, \ddot\cup\, D.LT$) can replace the multi-modal scan followed by a traditional join ($T \Join \ddot Scan(D.LT)$) or union ($T \cup \ddot Scan(D.LT)$).
As such, they also extract values for latent attributes from text but can leverage additional context given by their tabular operand for better and faster extractions.
For \singlerow{} latent tables, joins and unions are implemented as explained in Section \ref{sec:model}, feeding sequences as shown in Figure \ref{fig:model} \circled{1} into the model and then using the RD head for joins, and the AD head for unions for value extraction.

Joins and unions for \multirow{} latent tables are implemented using Algorithm \ref{alg:complex}.
The only difference is the additional context fed into the encoder (see lines 2 and 6 in Algorithm \ref{alg:complex}).
For joins, we feed additional table values added to each input sequence, coming from a tuple linked to the text of the sequence (as shown in Figure \ref{fig:model} \circled{1}).
For unions, we again randomly sample two rows from the tabular operand and feed these as example values the \method{}-model, which can be used by vertical self-attention to improve extractions.}

The join comes with opportunities for optimization as shown in Figure \ref{fig:join-cases}.
In the case \circled{1}, not all texts in the text collection may have a join partner in the table, especially if the table has been filtered beforehand (e.g., only patients with age < 18 are selected, and thus, only texts of such patients need to be processed).
As such, the path in the table data acts as an index to the text collection to decide which subset of text documents need to be scanned (i.e., we can avoid scanning all text documents).
For \multirow{} joins, there are two cases.
In Figure \ref{fig:join-cases} \circled{2} (a), we need to extract multiple diagnoses per text document for each tuple (i.e., patient) from the normal table.
Here, we must run Algorithm \ref{alg:complex} to first extract all diagnoses from the text.
Afterwards, the values for potential dependent columns (e.g., treatment) are extracted.
However, case \circled{2} (b) can be optimized thanks to the row-detect head, as shown below.
In this case, each text is about multiple patients and we have only one diagnosis per patient.
As such, multiple tuples from the table refer to the same text and we only need to extract a single row per patient tuple.
For this, we feed each text multiple times into the encoder, each time paired with a different patient tuple from the table.
The RD head is pre-trained to extract only values corresponding to the table tuple (i.e. patient) it is paired with, which allows us to extract multiple rows from a single text without Algorithm \ref{alg:complex}.
The difference to case (a) is that the tabular join partner already contains the values for the \idattr{} $la_{key}=name$.
Hence, we can skip extracting these in the first phase of Algorithm \ref{alg:complex}.
}

\begin{figure}[t]
  \centering
  \includegraphics[width=0.9\linewidth]{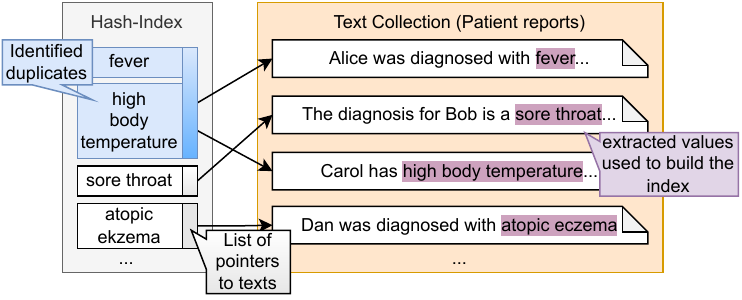}
  \caption{The multi-modal index used for selections. When extracting the values stored in the index from the texts (using attribute-detect), we use the duplicate-detect head to identify values that refer to the same concept. This allows the index to return texts of patients that have a fever when users query for patients with high body temperature.
  }
  \label{fig:index}
\end{figure}

{\subsection{Multi-Modal Selection}
\label{sec:selection}

The multi-modal selection $\ddot \sigma_{cond}$ filters data based on attributes extracted from text.
For example, users might be interested only in \emph{treatments} for patients diagnosed with \emph{fever}.
As discussed in Section \ref{sec:definitions}, the multi-modal selection comes in two variants.

\noindent\textbf{Variant 1.} The first variant $\sigma_{cond}(D.LT)$ takes a latent table as input and produces a (filtered) latent table as output (i.e., without materializing neither input nor output table).
Important is that the selection thus reduces the number of text documents in a collection before using them as input (e.g., to a join).
Moreover, the selection uses semantic similarity for filtering. In \method{}, we provide an implementation for this scan that uses a multi-modal index on the selection column.
}
The core idea of a multi-modal index is that it maps a search key for the query attribute (e.g., diagnosis) to text documents that contain the search key.
To construct a multi-modal index, our approach leverages the attribute-detect head to extract all values for the search key from all texts.
For building the index, we put the extracted values and the pointers to the text documents into a traditional index to be able to retrieve text documents quickly.
\method{} currently uses a hash index.
However, unlike traditional hash indexes, we group similar extracted values into one index entry using the duplicate-detect head of our \method{}-model, as depicted in Figure \ref{fig:index}.
This allows the index to return text documents containing the diagnosis of \emph{high temperature}, even if the user queries for \emph{fever}.

{
\noindent\textbf{Variant 2.} As discussed in Section \ref{sec:definitions}, \method{} comes with a second variant $\ddot\sigma_{cond}(T)$ that can be evaluated on a normal table.
Here, we might still want to select rows based on the semantic similarity of values extracted from the text (e.g., if the selection is executed after a multi-modal join).
Given a selection condition (e.g., diagnosis = fever), we embed it by feeding \texttt{<attribute> is <value>} (e.g., diagnosis is fever) into our encoder and compute a span-embedding of \texttt{<value>} as discussed before.
Then, we use the DD head of our model to decide whether a value embedding from the table and the selection value refer to the same semantic concept.
To support this selection, an important optimization is to keep the span embeddings of extractions created during a scan, join or union and attach it to the values of the respective output table. 
This avoids recomputing these embeddings if a selection follows one of those operators.
}

{
\subsection{Multi-Modal Aggregation}\label{sec:aggregation}

The last operation supported in \method{} is a multi-modal aggregation $\ddot \chi_{F, A'} (T)$ that can be used for group-by aggregation based on attributes extracted from text.
Like the second selection variant, this operation is designed to work on tables $T$ created from texts (e.g., to aggregate the result of the output of a multi-modal join).
Unlike a normal aggregation, the operation uses the semantic similarity of group-by values to form groups.
For example, assume we only use the patient reports as input and want to count how often a certain diagnosis was named across all texts.
The solution is first to extract all diagnoses using a scan and then use a multi-model aggregation to elegantly deal with cases where the same diagnosis is expressed differently in different texts (e.g., as fever and high body temperature) but is still counted as the same diagnosis.
More specifically, for the multi-modal aggregation, we compute a similarity matrix of the group-by values (as in line 2 in Algorithm \ref{alg:decoder:duplicate_detect}).
Afterwards, we use agglomerative clustering with a distance threshold to find all clusters (i.e., groups) for group-by (as in line 4 in Algorithm \ref{alg:decoder:duplicate_detect}).
Note that computing the distance matrix is quadratic in the number of input values, which becomes expensive for very high numbers of table rows.
Hence, in the future, we aim to replace the clustering-based aggregation algorithm with latent-semantic-hashing-based DBSCAN \cite{lsh-dbscan}, which is linear in the number of input embeddings.

}
\subsection{A New Pre-Training Corpus}\label{sec:corpus}

\noindent Unfortunately, currently no pre-training corpus exists that allows us to pre-train our \method{}-model as outlined in Section \ref{sec:model}.
In contrast to existing corpora such as \cite{book-corpus,webtables,gittables}, we need a different parallel corpus where the texts contain the same information (e.g., same entities) as the tables, and where we know which cell values also occur in a text, allowing us to mask them for pre-training.
Hence, we constructed a new parallel open-domain pre-training corpus from Wikidata and Wikipedia for pre-training multi-modal database models.
We open-source the corpus together with our code\footnote{\url{https://github.com/DataManagementLab/eleet}\label{footnote:github}}.

\begin{figure*}[t]
  \centering
  \includegraphics[width=\linewidth]{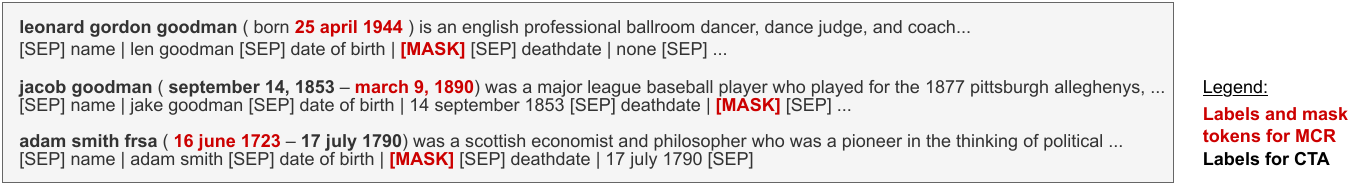}
  \vspace{-3.5ex}
  \caption{Example pre-training sample consisting of three rows and texts from our pre-training data set.
  }
  \label{fig:corpus-example}
\end{figure*}

The main idea of our data set is that we make use of T-REx \cite{trex}, a large-scale alignment of Wikidata triples and Wikipedia abstracts.
The T-REx data set contains 11 million alignments of Wikidata triples to Wikipedia abstracts.
All the 3.09 million abstracts occurring in T-REx are also part of our data set.
T-REx itself is created automatically using the distant supervision assumption for computing the alignment and can thus be noisy sometimes, but it allows us to construct a large pre-training corpus with objectives aligned to the downstream task of multi-modal database operations.
It has been used by other researchers for training their ML models \cite{kilt}.

Hence, we use the alignment of T-REx as a starting point to construct our parallel corpus of texts and relational tables.
T-REx contains information about the location of mentioned entities in the texts, which we can use as labels for our pre-training objectives.
As texts, we simply use the aligned Wikipedia abstracts and construct additional tables using Wikidata, by grouping similar entities together in a table and using Wikidata properties as columns.
We use several techniques to obtain a rich and diverse data set, e.g. we randomize column names using Wikidata's aliases or concatenate multiple abstracts to simulate texts about multiple entities.
See Figure \ref{fig:corpus-example} for an example training sample for pre-training.
In total, our pre-training data set consists of 8.2m such samples in its training set and 7.8k in its development set.
\section{Experimental Evaluation}
\label{sec:eval}

In this section, we present the results of our experimental evaluation, which justifies the design of \method{}.
To do so, we constructed a challenging benchmark containing 70 multi-modal query plans over four data sets that we publish along with this paper\textsuperscript{\ref{footnote:github}}.

\subsection{Evaluation Setup} \label{sec:eval_setup}

\noindent\textbf{Data sets.} Our benchmark consists of four data sets.
Based on each data set, we build a database consisting of 1-6 relational tables and 1-3 text collections; see Table \ref{tbl:benchmark} (upper part) for detailed statistics.
Note that all data sets are from different domains and include data that \method{} has not seen during pre-training.

(1) \emph{Rotowire}: The \emph{rotowire} data set \cite{rotowire} contains a text collection of 4850 basketball game reports with an emphasis on game statistics.
Hence, the values to be extracted are primarily numeric.
We complement the text collection with several tables of Basketball Players and Teams to enable multi-modal queries.
In total, this data set has 6 structured tables and 1 text collection, while 2 latent tables are derived from the text collection.

(2) \emph{T-REx (unseen):} The second multi-modal database is built using T-REx \cite{trex}.
Importantly, this data set is composed of three unseen domains that were not used in our pre-training: \texttt{nobel} prize winners, \texttt{skyscrapers}, and \texttt{countries}.
The data set uses Wikipedia abstracts as text collections and table rows are constructed from Wikidata.
In total, this data set has 6 structured tables and 3 text collections, while 6 latent tables are derived from the text collections.
Importantly, T-REx is constructed automatically and is thus too noisy to be used for evaluation.
Hence, we fine-tune the models on the three mentioned domains but evaluate only on queries from the \texttt{nobel} domain, which is the least noisy.

(3) \emph{Aviation}: Based on the \emph{aviation} data set from \cite{wannadb}, we construct a document collection, where each document is an aviation accident report published by the \href{https://www.ntsb.gov/investigations/AccidentReports/Pages/Reports.aspx?mode=Aviation}{United States National Transportation Safety Board}.
Attributes that can be extracted from the texts are the location of the accident, the severity of the damage, and so on.
In total, this data set has 3 structured tables and 1 text collection, while 1 latent table is derived from the text collection.

(4) \emph{Corona}: The final data set is again based on data used in \cite{wannadb}. It consists of one text collection containing the daily status reports by the German RKI.  From these texts, information like the number of persons infected by or recovered from Covid-19 can be extracted. We pair these texts with a single table (used for multi-modal unions only).
In total, this data set has 1 structured table and 1 text collection, while 1 latent table for the multi-modal union with the same attributes can be derived from the text.

\begin{figure}[t]
\begin{table}[H]
\small
\begin{center}
\caption{\label{tbl:benchmark}Statistics of our benchmark. The lower part indicates how often each multi-modal operator is used in queries.}
    \begin{tabular}{ p{2.7cm} | c c c c}
\toprule
 \textbf{Data set} & \textbf{Rotowire} & \textbf{T-REx} & \textbf{Aviation} & \textbf{Corona} \\
 \midrule
 \#text collections & 1 & 3 & 1 & 1 \\
 \#tables & 6 & 6 & 3 & 1  \\
 \#latent tables & 2 & 6 & 1 & 1  \\
 \#queries & 28 & 12 & 15 & 15 \\
 \#attributes latent tables & 21, 14 & 8 & 7 & 7 \\
 \#texts (test set) & 728 & 221 & 30 & 30 \\
 \midrule
 \#join & 10 & 2 & 5 & 0 \\
 \#union & 4 & 6 & 5 & 15 \\
 \#scan & 14 & 4 & 5 & 0 \\
 \#selection & 6 & 2 & 3 & 0 \\
 \#aggregation & 6 & 2 & 3 & 0 \\
 \bottomrule
\end{tabular}
\end{center}
\end{table}
\end{figure}

\begin{figure*}
    \centering 
    \includegraphics[width=\textwidth]{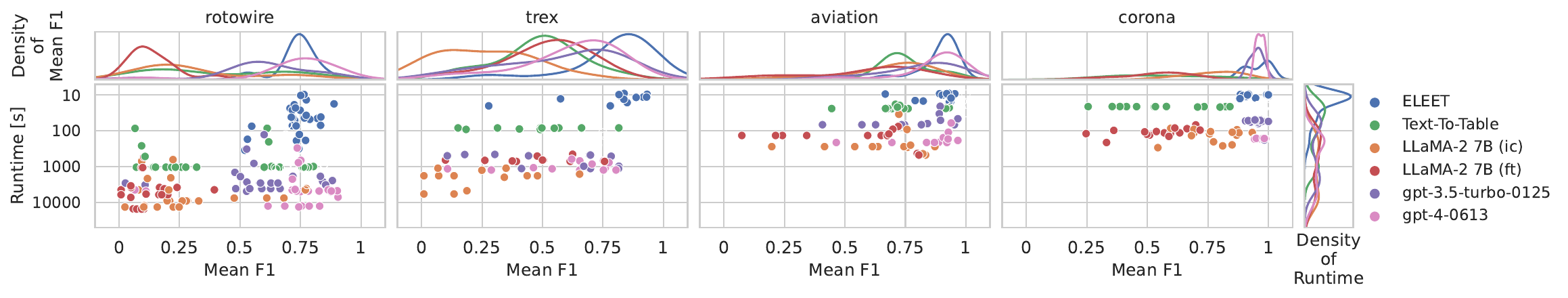}
    \caption{
        We plot the runtime and F1 scores of each approach on all 70 queries in our benchmark in a scatter plot.
        Due to the vastly different runtimes of individual queries, the runtime is shown in log scale.
        The GPT-models and LLaMA (ic) use few-shot prompting (in-context learning) and the other models are fine-tuned.
        We see that \method{} is always the fastest approach, while the baselines are sometimes several orders of magnitude slower.
        The density plot above the scatter plot makes it easy to see that \method{} is among the most accurate approaches for all data sets despite being a small model.}
    \label{fig:exp1}
\end{figure*}

\noindent\textbf{Query Generation.} Based on these data sets, we generate 70 query plans with 1-3 multi-modal operators and 0-1 traditional operators. See Table \ref{tbl:benchmark}  (lower part) for statistics of the generated queries.
{
A list of all queries is available in appendix \ref{sec:benchmark}.
}

\noindent\textbf{Baselines.}
We compare the \method{} to several strong LLM and SLM baselines.
Since there are no other systems so far that do joins and unions on multi-modal table/text data, we build our baselines from recent state-of-the-art models from the NLP community.

(1) \emph{Text-To-Table \cite{text-to-table}}: Text-To-Table is a machine learning model based on BART (-base) \cite{bart} with a similar size as our model that can be trained to translate texts to tables.
Different from the \method{}-model, it did not receive any special pre-training (i.e., it needs to be trained with the BART-weights as starting point for each new data set).
Moreover, it uses an autoregressive decoder, which is less efficient than our lightweight decoder.

(2) \emph{LLaMA-2 (7B) \cite{touvronLlamaOpenFoundation}}: LLMs such as LLaMA can also be used to translate texts to tables using few-shot prompting (i.e., in-context learning, ic) \cite{gpt3, weiEmergentAbilitiesLarge2022} or fine-tuning (ft).
{
We evaluated several prompts and found the following prompt to work well: ``Transform the input text into a <name of latent table> table. $\hookleftarrow$ Only output the table in CSV-Format without explanations. If there is no information for a cell, leave it empty. The header row is: <columns of the result table>. $\hookleftarrow$ Input: <text to translate to a table>''
}
For in-context learning, we add as many task demonstrations as the context window of 4096 tokens allows.
For fine-tuning, we use Q-LoRA \cite{qlora}.

(3) \emph{GPT-3.5 (gpt 3.5 turbo) \cite{gpt3} and GPT-4 (gpt-4-0613) \cite{openaiGPT4TechnicalReport2023a}}: GPT-3.5 and GPT-4 are even larger LLMs with 175 billion parameters and 1.76 trillion parameters respectively.
{
We use few-shot prompting for both models, using the same prompts as for LLaMA.
}
GPT-4 is used by Evaporate-Direct \cite{aroraLanguageModelsEnable2023b} to extract information from semi-structured documents (e.g., XML-Documents) similar to our baseline.
Unfortunately, all OpenAI models (GPT-3.5 and GPT-4) are closed source and thus can only be accessed via the API and not deployed on our hardware for comparing runtimes.
Nonetheless, we compare in Experiment 1 against these models to get a basic understanding of their accuracy.
However, since these models use many orders of magnitude more parameters than \method{} and thus are naturally much slower,  we skip these baselines in Experiments 2 and 3 and concentrate on the smaller language models such as LLaMA-2 and investigate how fine-tuning for those smaller models helps to achieve better accuracies compared to \method{}.

We do not compare against WannaDB \cite{wannadb}, because it requires user interaction for information extraction.
Moreover, we do not compare against Evaporate-Code \cite{aroraLanguageModelsEnable2023b}, because it is designed to extract information from semi-structured documents (e.g., XML, JSON) and not continuous text.
To run MMOps with the baselines, we use their capability to transform texts to tables to transform all documents in the document collection to tables.
Afterwards, we perform the corresponding traditional database operation.

\noindent\textbf{Training and Fine-tuning.}
All experiments were executed on a \href{https://images.nvidia.com/aem-dam/en-zz/Solutions/data-center/dgx-a100/dgxa100-system-architecture-white-paper.pdf}{DGX A100 server}.
For pre-training, we used 4 GPUs, which took approximately 8 days to train our model for 6 epochs on our pre-training data set.
For fine-tuning and inference --- in particular, for the performance measurements --- we used 2 GPUs only (for all models except GPT-3/4, which we cannot run on our hardware).

\noindent\textbf{Metrics.}
In our experiments, we focus on two main dimensions:
(1) First, we measure the quality of the query results computed by \method{} compared to the baselines.
{
We use Exact Match (EM) to evaluate the quality of query results.
All datasets come with a ground truth translation for all texts in the dataset.
A value is considered correct if it exactly matches its label.
We compute an F1 score for each text and report their mean.
While extractive models will generally output the values exactly how they are mentioned in the text (e.g., US president), generative models might output the values in arbitrary form (e.g., President of the United States).
Hence, the datasets come with different alternatives for each value, each of which counts as correct.
}
For aggregations, we group-by a certain attribute and collect for all other attributes the values for all groups.
We then compute an F1 score per group and report the mean.
(2) To compare the efficiency, we compare the runtime for each query.

\subsection{Exp. 1: Runtime vs. Accuracy}
Our main goal is to show that \method{} is orders of magnitude more runtime efficient than the SLM and LLM baselines while being more accurate.
We therefore execute the full set of queries on all data sets using \method{} and all baselines.
Since the collection of training data per text collection is expensive, we focus on a scenario where only limited data is available for fine-tuning.
Hence, in this experiment, we limit the number to 64 labeled texts for fine-tuning per data set (for \method{}, Text-To-Table, LLaMA).
For LLaMA, GPT-3, and GPT-4, we use few-shot prompting and include the annotated examples in the input prompt, as explained before.
We think this is a reasonable number of texts that can be labeled manually.
We later evaluate in more detail how a different amount of labeled training data for fine-tuning affects the accuracy of \method{} compared to the baselines.

\noindent\textbf{Overall Results.}  In Figure \ref{fig:exp1} (lower part) we plot runtime and F1 scores for each query in our benchmark in a scatter plot. 
Overall, we see that \method{} provides high accuracy (i.e., a high F1-score) while being fast in execution (i.e., often in the order of seconds).

\noindent\textbf{Runtime.} As Figure \ref{fig:exp1} shows, \method{} is the fastest approach, while other approaches are up to $575\times$ slower. 
For instance, the following query which is included in our testing set (i.e.,
$player\_info$
$\Join$
$player\_to\_reports$
$\ddot \Join$
$reports$
$.player$)
takes about three minutes with \method{}, 17 minutes using Text-To-Table, 2.5 hours using LLaMA, 1.5 hours using GPT-3, and almost 4 hours using GPT-4.
This shows how expensive in terms of runtime LLMs are when applied to many texts.
Overall, \method{} clearly outperforms the LLM-based baselines (LLaMA, GPT-3 and GPT-4), even though the GPT-models run on the hardware of OpenAI.
But even the comparatively small Text-To-Table is significantly slower than \method{}, which we attribute to its use of a transformer-based autoregressive decoder that requires many passes through the model compared to our model.

\noindent\textbf{Accuracy.} 
The density of F1 values (upper part of Figure \ref{fig:exp1}) nicely shows that \method{} usually exceeds the performance of the baselines despite being a small model.
For aviation and corona, \method{} achieves F1-scores of over 90\% for most queries. 
For the other two data sets, \emph{rotowire} and \emph{T-REx}, which are more challenging, the F1 scores of \method{} cluster above and around 75\% and 80\%, outperforming the baselines.
The most competitive model is GPT-4, which is very accurate on aviation and corona.
However, due to its immense size, it is much computationally more expensive as discussed before.
Moreover, we see the effect of our pre-training when comparing the performance to Text-To-Table and fine-tuned LLaMA, which have F1 scores of around 25\% for most queries on \emph{rotowire}.
64 samples are not enough for these models to allow for decent extractions.

\subsection{Exp. 2: Varying Data Sizes for Fine-tuning}

In the previous experiment, we have seen that \method{} can accurately compute multi-modal queries with only a few fine-tuning samples.
However, users might often have different requirements on the quality of query results and access to different amounts of labeled texts for training.
In this experiment, we show how \method{} behaves with training sets of various sizes compared to the baselines.
Here, we use five queries on the \emph{rotowire} data set for testing. The queries used in this experiment cover all different MMOps:

\begin{itemize}
    \item \textbf{Scan:} $\ddot Scan(reports.player)$
    \item \textbf{Join:} $(player\_info \Join player\_to\_reports) \ddot \Join reports.player$
    \item \textbf{Union:} $player\_stats \; \ddot \cup \; reports.player$
    \item \textbf{Selection:} $\ddot Scan(\ddot\sigma_{Points=28}(reports.player))$
    \item \textbf{Aggregation:} $\ddot \chi_{name}(\ddot Scan(reports.player))$
\end{itemize}

We fine-tune several models for \method{}, Text-To-Table, and LLaMA using a varying number of labeled texts.
We vary between 4 labeled texts up to the full training set (3398 texts) and report the mean F1 score of all queries from above for each model.
As discussed before, we limit ourselves to models that run on our hardware.

\begin{figure}
    \centering 
    \includegraphics[width=\columnwidth]{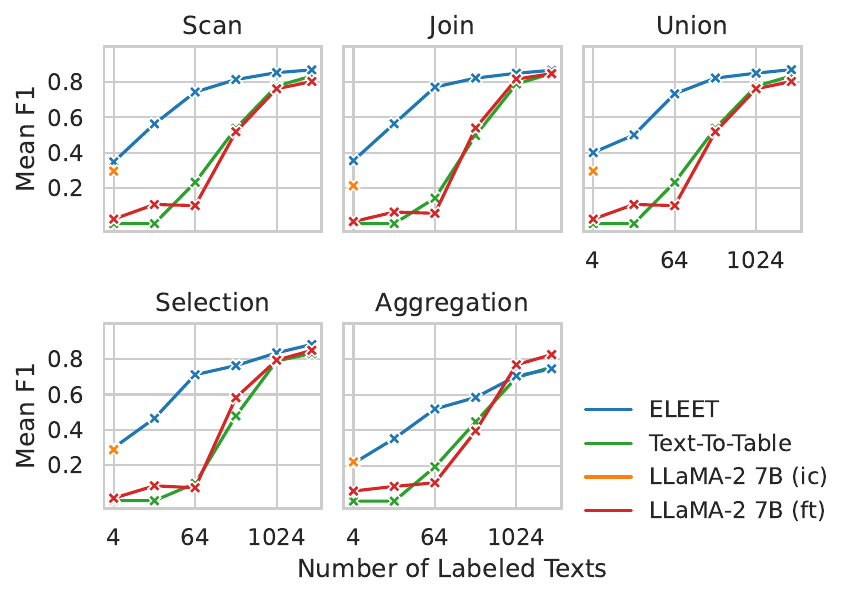}
    \caption{Mean F1 scores for different queries and training set sizes on \emph{\emph{rotowire}}.
    The F1 scores for all models that use fine-tuning increase as the number of labeled texts for training increases, but \method{} consistently outperforms all baselines.
    }
    \label{fig:exp2}
\end{figure}

\noindent\textbf{Accuracy with varying training data.} 
Figure \ref{fig:exp2} shows the results.
With limited training data (4-16), only LLaMA using in-context-learning and \method{} achieve accuracies of around 40\%.
The other fine-tuned methods LLaMA and Text-To-Table struggle in these few-shot scenarios due to the missing specialized pre-training.
When we increase the amount of training data for fine-tuning, the accuracies of all fine-tuned methods increase.
LLaMA using in-context learning, on the other hand, cannot use this additional training data due to the limited context size.
Overall, we see that \method{} outperforms both fine-tuned baselines across all training set sizes.
In particular, for scans, joins, and unions, \method{} achieves a better Mean F1 score than Text-To-Table and LLaMA even when trained on the entire data set of \emph{rotowire}.
We achieve an unmatched F1 score of 87\% for joins, unions, and scans when trained on the entire data set.

\subsection{Exp. 3: Runtime of Multi-modal Operators}

\begin{figure}
    \includegraphics[width=\columnwidth]{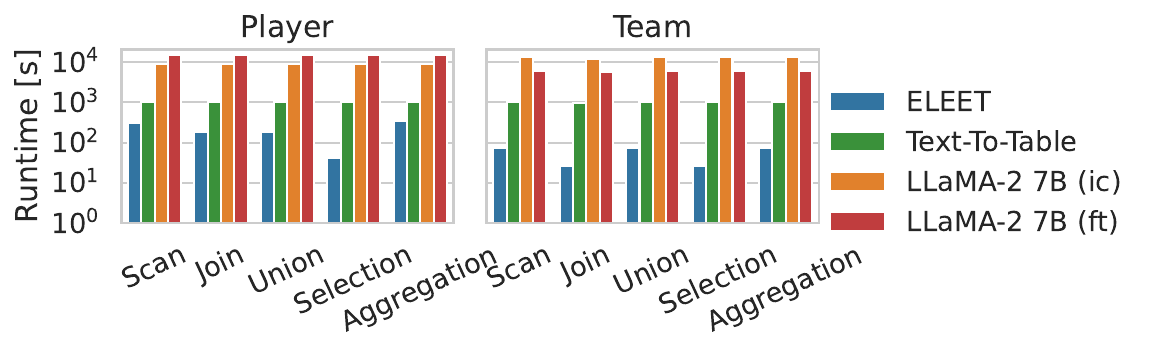}
    \caption{Runtime comparisons on the two latent tables of rotowire (\texttt{reports.player} and \texttt{reports.team}, containing Player and Team statistics).
    Comparing the query runtime of \method{} on the different queries, we see that joins are faster than scans or unions as they do not run Algorithm \ref{alg:complex}.}
    \label{fig:runtime}
\end{figure}

In the next experiment, we zoom into query runtime and show what typical query runtimes for \method{} are, and how efficient each individual operator is.
Figure \ref{fig:runtime} shows the results.

\textbf{Overall results.}
As we have seen before, \method{} vastly outperforms all baselines in terms of query runtime.
However, comparing the query runtime of different queries using \method{}, we see big differences.
In particular, joins are more efficient than unions or scans, because they do not necessitate execution of Algorithm \ref{alg:complex} as discussed in Section \ref{sec:join_union}.
The multi-modal union and scan will extract multiple tuples per game report by first extracting the name of the Team/Player (name is the \idattr{} of the latent \texttt{reports.player} and \texttt{reports.team} tables).
Only in a second iteration are the statistics of each Player or Team extracted.
The join on the other hand uses the signal from the table, the document collection is joined with (\texttt{player\_info} and \texttt{team\_info}).
These already contain names and other information of players and teams, and hence the first iteration can be skipped.
This effect can be best seen in Figure \ref{fig:runtime} (right), where a join with the latent \texttt{teams} table takes 26 seconds, while scans and unions take 70 seconds.
Moreover, selection operations can reduce query runtime to the order of seconds.
This is due to the use of indexes, allowing efficient execution if only a few texts need to be processed based on the filter predicate.
A similar effect also holds for selective join, as we show next.

\subsubsection{Selective Multi-modal Join Queries}

To investigate the scenarios of selective joins where the data in the tabular input helps to reduce the number of texts we need to analyze, we look at such join queries on the latent player table of \texttt{rotowire}:

\begin{equation*}
    (\sigma_{cond}(player\_info) \Join player\_to\_reports) \ddot \Join reports.player
\end{equation*}

In these queries, $cond$ is an arbitrary condition that selects a certain amount of players.
The join queries in this experiment have different selectivities; i.e., the query only selects a few \texttt{players} before executing the multi-modal join with the text collection.
Since tuples in the table are linked to the game reports (e.g., a tuple about a player is linked to all the game reports the player participated in), this usually results in only a few selected texts as well (i.e., the filter on the table acts as an index into the text collection).

\noindent\textbf{Runtime for different selectivities.} Figure \ref{fig:selective-runtime} (left) shows how the different selectivities affect the overall query runtime of the queries containing the selective multi-modal join operator.
Here we encounter another benefit of \method{}: Since \method{} only needs to materialize those rows that have a join partner in the tabular operand (by feeding table tuples in the model), the runtime is reduced to the order of a few seconds.
All other approaches translate the entire text to a table, which results in runtime overhead.

\subsubsection{Multi-Modal Selection}

In the second scenario, we analyze the runtime of simple queries that only need to scan a subset of texts in a text collection using a multi-modal selection operation (variant 1) on extracted attributes (e.g., $\ddot Scan(\ddot\sigma_{Points=28}(reports.player))$).
If implemented naively, this query results in a costly operation since, independent of the selectivity, all texts need to be first transformed to tuples to judge which texts qualify for the filter condition.
Instead, \method{} uses an index, as explained in Section \ref{sec:selection}.

\begin{figure}
    \centering 
    \includegraphics[width=\columnwidth]{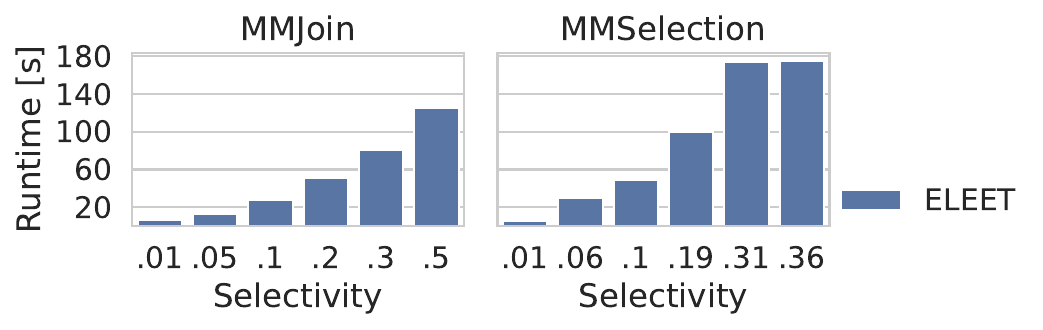}
    \caption{
      Runtime behavior of \method{} for selective queries on rotowire.
      (Left) The query selects a subset of rows (i.e. players) from the table before a multi-modal join, which results in fewer texts being processed.
      (Right) A multi-modal selection operation (with subsequent scan) using our secondary index.
      The index reduces the number of processed texts, allowing much faster runtimes for lower selectivities.}
    \label{fig:selective-runtime}
\end{figure}

\noindent\textbf{Runtime for different selectivities.}
Figure \ref{fig:selective-runtime} (right) shows the runtime with different filter conditions resulting in different amounts of selected texts.
For queries with low selectivity, \method{} can again compute the query results in a few seconds.
The naive solution to translate all texts into tables first and then doing the selection afterwards would always take a few minutes, independent of selectivity.

\subsection{Exp. 4: Ablation Study for Pre-Training}\label{sec:ablation}

In the next experiment, we show the effect of our pre-training objectives by comparing it to other existing pre-training procedures.

\noindent\textbf{Alternative pre-training procedures.}
{
To show the importance of multi-modal pre-training that teaches the model the necessary skills to perform MMOps, we examine whether our pre-training procedure results in better extractions compared to other pre-training procedures.
}
We compare against two alternative pre-training procedures: the pre-training procedure used for BERT \cite{bert} and the pre-training procedure used for TaBERT \cite{tabert}.
The BERT model aims at natural language understanding and is thus pre-trained on plain text only, meaning it has never seen any tabular data before.
TaBERT, on the other hand, is also pre-trained on a parallel corpus of texts and tables scraped from the web.
However, in the parallel corpus of TaBERT, texts and tables are often only vaguely related.
Moreover, the pre-training objectives are designed for tasks like text-to-SQL and are not ideal for enabling MMOps.

\begin{figure}
    \centering
    \includegraphics[width=\columnwidth]{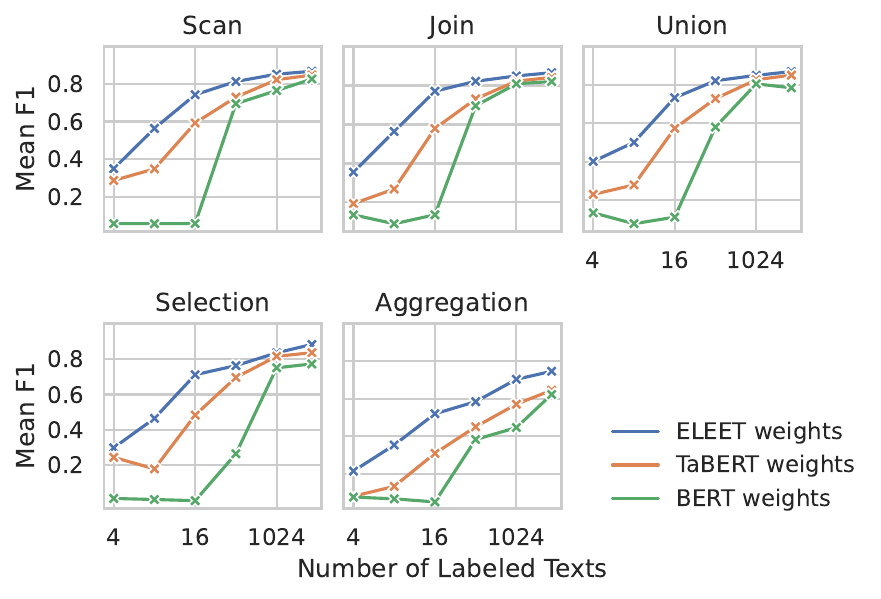}
    \caption{Comparison of the query result quality when using \method{} with different pre-trained weights.
    The pre-trained weights resulting from our pre-training procedure result in better extractions across training set sizes.}
    \label{fig:ablation}
\end{figure}

\noindent\textbf{Effect of different pre-training procedures.}
Figure \ref{fig:ablation} shows how the different pre-training procedures affect the quality of query results when varying the number of samples for fine-tuning.
{Especially with only limited fine-tuning data, our model can consistently extract tabular data from text more accurately.
Comparing TaBERT to BERT, we see that TaBERT can consistently outperform BERT since it is already using a form of tabular pre-training.}

\vspace{6ex}
\subsection{Exp. 5: Architectural Design Decisions} \label{sec:ablation-model}

{

In this section we justify the design decision of our model, in particular the use of an extractive model as well as the use of vertical-self attention in our model.

\noindent\textbf{Extractive vs. Generative Models.}
As explained in Section \ref{sec:design_principles}, we chose an extractive decoder.
The alternative, a generative auto-regressive decoder generates the output tokens one-by-on and thus requires many passes through the model.
This can also be seen in our previous experiments.
In Figure \ref{fig:exp1}, we show that \method{} is much faster than Text-to-Table \cite{text-to-table} across all queries.
Text-To-Table is an SLM with a similar number of parameters as \method{} that uses autoregressive decoding.
This shows that for the task of transforming texts to tables, extractive decoding is much faster than autoregressive decoding.

Moreover, to isolate the effect of extractive decoding versus generative decoding using an autoregressive model, we did a micro-benchmark, where we stripped down our model to a simplified version,
essentially consisting only of a transformer-based encoder and either an extractive decoder or a transformer-based autoregressive decoder.
Hence, this stripped-down model can be used to either extract values from text (as we do) or generate them using its autoregressive decoder (as the GPT models do).
For the autoregressive model, we use 12 transformer layers for the Encoder (as discussed in Section \ref{sec:model})
and another 12 for the decoder.
To make the comparison fair, we increased the number of encoder layers of the extractive model to 24. 
Hence, both variants have the same number of transformer layers and are equivalent in size.
For the evaluation, we used the SQuAD \cite{squad} dataset, which requires extracting a single value from each text (can be seen as a table with a single column) to avoid the effects of different output serializations for the generative model.

Figure \ref{microbenchmark} shows that the extractive variant is 25 percent faster than the generative variant, which means the absolute runtime difference scales linearly with the number of documents in the document collection.
Moreover, this effect is more pronounced when the output sequences are longer, i.e., when the output is not just a single value consisting of a few tokens.

\begin{figure}
    \centering
    \includegraphics[width=0.6\columnwidth] {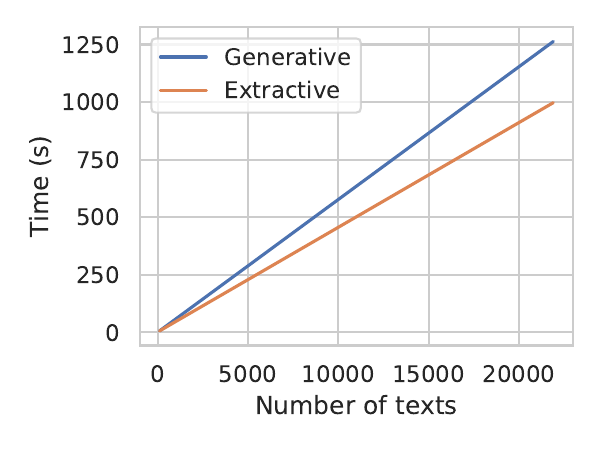}
    \caption{Comparison of extractive and generative decoding on the SQuAD dataset.
    The generative variant needs 25 percent more time than the extractive variant.
    }
    \label{microbenchmark}
\end{figure}

\noindent\textbf{Vertical Self-Attention.} As explained in Section \ref{sec:model}, one crucial component of our model is the use of vertical self-attention \cite{tabert} to let signal flow between input sequences.
Especially for unions, letting signal flow between the different input sequences is important.
In a union, the other rows contain example values from the tabular operand that can be used as additional context that can help during extraction.
These example values can, for example, resolve potential ambiguities in the names of the query attributes.

To evaluate this effect in isolation, we created a new text collection of ambiguous patient reports that contain two kinds of diagnoses:
First, the patient reports contain diagnoses regarding the patients' health problems.
Additionally, technical equipment (e.g., the equipment in the room) is also broken, and the text contains diagnoses regarding their malfunctioning.
We constructed the dataset using a set of patient report templates and letting GPT-4 fill in the blank health-related and technical diagnoses.

Hence, in a latent table with columns name and diagnosis, it is unclear which diagnosis should be extracted, the health-related one or the technical equipment-related one.
However, this ambiguity can be resolved when performing a union with either a health-related diagnosis table or a
technical-equipment-related diagnosis table, if the model is able to let the signal flow from the example rows to the row containing the mask tokens for extraction.
Table \ref{tab:vertical-attention} shows that \method{} is able to resolve this ambiguity when using the example rows as additional context.
With vertical self-attention, \method{} is able to extract the correct diagnosis almost perfectly.
Without vertical self-attention, the aforementioned ambiguity cannot be resolved, causing the F1-score to drop to around 50\%.

}

{

\begin{table}[h]
    \centering
    \caption{Comparison of \method{} with and without vertical attention on a synthetic dataset with ambiguous column names.
    Vertical attention allows signal flow between the input rows, which helps resolve ambiguous column names.}
    \begin{tabular}{l|c|c}
        Model & user's interest & mean F1 \\
        \hline
        \method{} & technical diagnosis & 0.99 \\
        \method{} & health diagnosis & 0.99 \\
        \method{} (w/o vertical attention) & technical diagnosis & 0.41 \\
        \method{} (w/o vertical attention) & health diagnosis & 0.55 \\
    \end{tabular}
    \label{tab:vertical-attention}
\end{table}

\subsection{Exp. 6: Automatic Latent Table Registration} \label{sec:auto-registration}
As explained in Section \ref{sec:data-model}, registering a latent table comes with some manual overhead.
More specifically, in order to register a latent table, users have to define a schema, define whether a latent table is \singlerow{} or \multirow{} and in the latter case, specify the \idattr{}.
To reduce the manual overhead, we tried to automate the selection of the \idattr{} and the decision of whether a latent table is \multirow{} such that the user does not need to make this decision by screening texts manually.
We used Chat-GPT-3.5 and prompted it to decide what the \idattr{} of a latent table should be and whether a latent table is \multirow{} (see prompt below).
In the prompt, we put a small sample of three example texts and the schema of the latent table.
Interestingly, we found that Chat-GPT-3.5 (gpt-3.5-turbo-0125) was able to correctly identify the \idattr{} and whether a latent table is \multirow{} for all latent tables of our four datasets as shown in Table \ref{tab:chat-gpt}.
As this process needs to be done only once per latent table (and not per document) and latent table registration happens before query execution time, we believe it is suitable to use a large language model like Chat-GPT-3.5 for this task.

\textbf{Details on prompt.} We used the following prompt for the identifying attribute:

\vspace{0.5em}
\texttt{
    I want to transform the information stored in texts into a table. The table should have the following columns: <columns>. $\hookleftarrow{}$
In a first step, please specify \textbf{one of the columns that act as a document-level key, meaning that all extracted values should be unique per document}.
Only output the name of that column without any explanations. $\hookleftarrow{}\hookleftarrow{}$
Sample of texts: $\hookleftarrow{}$
<sample of three texts>
}
\vspace{1em}

We used the following prompt for classifying latent tables on multi-row texts:

\vspace{0.5em}
\texttt{
I want to transform the information stored in texts into a table. The table should have the following columns: <columns>. $\hookleftarrow{}$
In a first step, please \textbf{determine if one or many rows are extracted per text document. Please only output "many" or "one"}
without any explanations.  $\hookleftarrow{}\hookleftarrow{}$
Sample of texts: $\hookleftarrow{}$
<sample of three texts>
}
\vspace{1em}

\begin{table}
    \centering
    \caption{Chat-GPT-3.5 is able to correctly identify the \idattr{} and whether a a latent table is \multirow{} on all of our four datasets.}
    \begin{tabular}{l|c}
        Task & Accuracy \\
        \hline
        Identify \idattr{} per table & 100\% \\
        Identify \multirow{} per table & 100\% \\
    \end{tabular}
    \label{tab:chat-gpt}
\end{table}

\subsection{Exp. 7: Varying Text Lengths} \label{sec:breakdown-textlength}

\begin{figure*}
    \centering
    \includegraphics[width=0.75\textwidth] {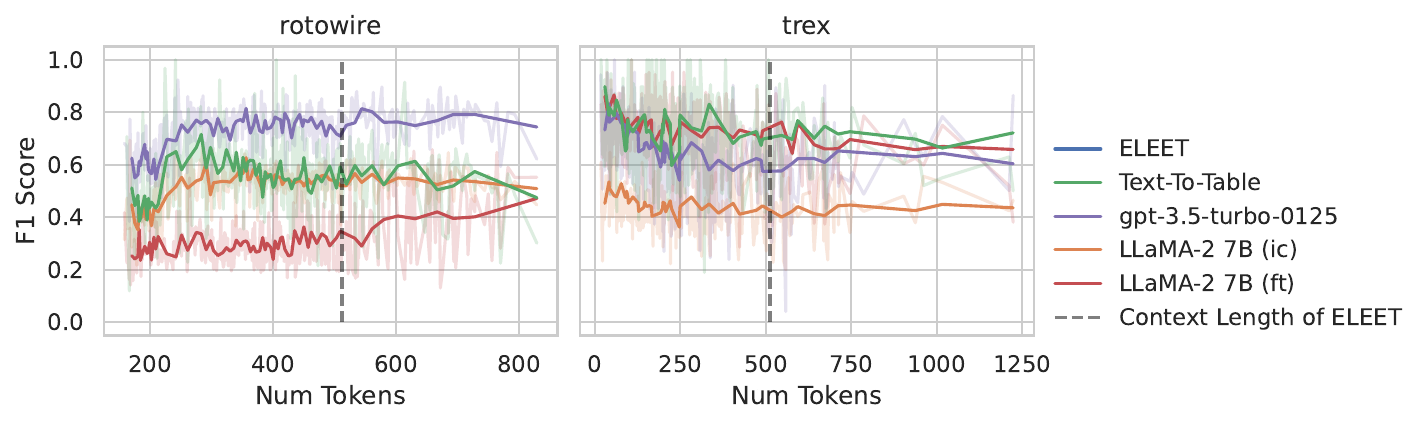}
    \caption{Extraction quality of \method{} broken down by length of text.
    The dashed black line denotes the context length of the  \method{}-model (512 tokens). 
    The performance of \method{} is stable under increasing text length.
    Important is that \method{} has the best F1 score despite having a smaller context window than the baselines and the accuracy does not drop for texts that are longer than the context length of the \method{}-model (right to the dashed line).
    }
    \label{fig:exp1-text-length}
\end{figure*}

Our work focuses on small to medium-sized texts (such as patient reports) but can also support longer texts using a windowed approach as explained in Section \ref{sec:scan}.
The main idea is to extract the values from longer text by processing them using a sliding window.
We can feed each text window independently into the \method{}-model and collect the extracted values across windows afterwards to compose the output table.
For \multirow{} texts, the procedure is slightly more involved due to the two phases of Algorithm \ref{alg:complex}.
Here, we again process the different text windows independently for each phase.
However, after extracting the values for the \idattr{} in the first phase, we need to first collect and deduplicate all extracted values \emph{across windows}, before we can continue with the second phase.
After extracting all values for all text windows in the second phase, we can collect them to compose the output table.

We found that this windowed approach works well and does not come with a degradation in extraction quality, as shown below.
This indicates that large context windows are not required to extract values from text.
To show that our model works well independent of text length, we break down the extraction quality by text length on the rotowire and T-REx datasets, which are the two datasets that contain long texts.
For this experiment, we removed all aggregations and selections from our benchmarks to show the effects of pure extractions.
Then, we executed all queries on the datasets and recorded the results.
For each tuple in an output table, we trace back from which text the tuple is coming (i.e., from which text have the tuple's values been extracted).
Then, we compute an F1 score for each text separately by only considering the tuples that were produced from each text.
Figure \ref{fig:exp1} shows the result, where we plot the text length on the X-axis and the F1 score for each text on the Y-axis.
The lines in the background show the F1 scores of each text, while the lines in the foreground are a smoothed version of the results by averaging the over the last 10 texts.
The dashed black line denotes the context length of the  \method{}-model (512 tokens). 
We can clearly see that the performance of \method{} is stable under increasing text length.
Important is that \method{} has the best F1 score despite having a smaller context length than the baselines, and the accuracy does not drop for texts that are longer than the context length of the \method{}-model (right to the dashed line).
}
\section{Related Work}\label{sec:related-work}
\noindent\textbf{Multi-Modal Data Systems.}
{
Integrating textual data into data systems is a long-standing problem.
Early work implemented a join that retrieves the most relevant documents for each tuple without extracting any structured data from the text \cite{chaudhuriJoinQueriesExternal1995}.
Later systems allowed users to write small extraction functions or UDFs to extract structured data from text and then allowed them to easily combine these extraction functions by writing SQL or datalog queries \cite{datalog_ie, chuRelationalApproachIncrementally, dataTamer, deepdive}.
However, these systems still require the user to write extraction functions, which is not necessary in \method{}.
Other works focus only on filtering the multi-modal data using natural language filters \cite{thalamusdb, subjective-databases}.
Some early systems extract data from texts by using techniques such as information extraction, named entity recognition, part-of-speech tagging, and/or hand-designed grammars or rules
\cite{structuredTextSearch, structured_web_text_querying, tari2010incremental, systemt}.
However, they typically do not consider the multi-modal case where tabular data is available in addition to texts.
}
More recently, NeuralDB \cite{neuraldb} and WannaDB \cite{wannadb} use pre-trained language models to run queries directly on text documents, similar to \method{}.
Both systems, again, do not consider the multi-modal case and do not support \multirow{} operations, where multiple tuples are extracted from a single text document.
Symphony \cite{symphony} can query multi-modal data using natural language.
The setting in data lakes differs from databases since the main concern is retrieving data from multi-modal data sets.
For retrieval, they propose an information compression pre-training objective to embed many modalities in the same latent space.
Caesura \cite{urbanCAESURALanguageModels2023} uses LLMs to generate multi-modal query plans similar to ours.
Plugging \method{} into Caesura is an interesting avenue for future work.

\noindent\textbf{Extraction of Tabular Data.}
{
GIO \cite{gio} and Evaporate \cite{aroraLanguageModelsEnable2023b} tackle the related problem of translating custom data formats (e.g., machine logs) or semi-structured documents (e.g., XML), respectively, into tables using code generation.
While GIO uses template-based code generation, Evaporate uses LLMs such as GPT-4.
However, we found that code generation is hard with free-form text, as in our setting.
}
Text-to-Table \cite{text-to-table} and STable \cite{stable} are sequence-to-sequence models trained to transform tables into text.
Both introduce several model adjustments to ensure that the model outputs a correctly structured table.
Different from Text-To-Table, STable can output table cells in arbitrary order.
Unlike our work, both are trained in a supervised manner from scratch for every new data set.

\noindent\textbf{LLMs for Data Management.} By now, many research groups have integrated language models into data systems to tackle various data management tasks.
Language models have been used to tune databases \cite{trummerDBBERTDatabaseTuning2022a}, solve data engineering tasks like entity matching, entity resolution, or missing value imputation \cite{FM-wrangle, ditto, flexer, machop, unsupervised_matching_data_text},
or augment databases with knowledge stored inside of LLMs \cite{OmniscientDBLargeLanguage, query_llms_with_sql}.
Moreover, Foundation Models for data management promise to be a solution for many different data management problems \cite{tablegpt, vogelFoundationModelsRelational2023}.

 \noindent\textbf{Pre-training Models.} Large pre-trained language models \cite{elmo, bert, roberta, gpt3} are by now dominating NLP and are quickly adapted for multi-modal \cite{vilbert, vl-bert, lxmert, visual-bert} and tabular data \cite{turl, tabbie, tuta}.
 To reduce the overhead of adaption to downstream tasks, pre-training objectives began to be more aligned with the downstream task for many core-NLP \cite{sspt, marge, span-bert-a, span-bert-b} and also structure-aware tasks \cite{tapex, grappa, gap, mqa-qg}.
 Most similar to our pre-training objectives are those pre-training procedures that rely on weak or distant supervision to align pre-training more to the downstream task.
 ReasonBERT \cite{reason-bert} uses a pre-training objective inspired by distant supervision for the downstream task of multi-hop hybrid question answering.
 StruG's \cite{strug} pre-training data set designed for the text-to-SQL task is based on the table-to-text data set ToTTo \cite{totto}, which was extracted from Wikipedia using heuristics and is much smaller than our data set.

\balance{}

\section{Conclusions and Future Work}
\label{sec:conclusion}

We presented  \method{}, a new execution engine that allows users to seamlessly query textual and tabular data.
The cornerstone of \method{} is the concept of multi-modal database operators, which are realizable using a small pre-trained language model.
As a result, multi-modal queries containing multi-modal database operators can be executed on new data sets with only minimal fine-tuning overhead and high performance.
{
Clearly, there are still many challenges when integrating \method{} into a real database system.
The query parser must be able to instantiate the multi-modal query plans containing multi-modal and traditional operators.
The query optimizer must reason about the effects of replacing traditional operators with multi-modal ones.
}
Moreover, an extension to other modalities like images is also an interesting avenue for future work.



\begin{acks}
This research is supported by the Hochtief project AICO (AI in Construction), the German Federal Ministry of Education and Research (BMBF), and the state of Hesse through the NHR Program. This work was also partially funded by the LOEWE Spitzenprofessur of the state of Hesse. Additional support was provided by hessian.AI and DFKI Darmstadt. We are also grateful to Torsten Gallen and Kelvin Chui for the insightful discussions.
\end{acks}

\bibliographystyle{ACM-Reference-Format}
\bibliography{main}

\appendix
\clearpage
\onecolumn
\section{Benchmark Queries} 
\label{sec:benchmark}

\noindent In the following, we list all queries used as benchmarks for the evaluations in Section \ref{sec:eval} in the paper.

\subsection{Rotowire}
\begin{enumerate}
\item $(player\_info \, \Join_{name} \, player\_to\_reports) \, \ddot \Join_{path, name} \, reports.Player$
\item $player\_stats \, \ddot \cup \, reports.Player$
\item $\ddot Scan(reports.Player)$
\item $(player\_info \, \Join_{name} \, player\_to\_reports) \, \ddot \Join_{path, name} \, \ddot \pi_{name, Points, Assists, Steals}(reports.Player)$
\item $\pi_{name, Points, Assists, Steals}(player\_stats) \, \ddot \cup \, \ddot \pi_{name, Points, Assists, Steals}(reports.Player)$
\item $\ddot Scan(\ddot \pi_{name, Points, Assists, Steals}(reports.Player))$
\item $(\sigma_{0.01}(player\_info) \, \Join_{name} \, player\_to\_reports) \, \ddot \Join_{path, name} \, reports.Player$
\item $(\sigma_{0.05}(player\_info) \, \Join_{name} \, player\_to\_reports) \, \ddot \Join_{path, name} \, reports.Player$
\item $(\sigma_{0.1}(player\_info) \, \Join_{name} \, player\_to\_reports) \, \ddot \Join_{path, name} \, reports.Player$
\item $\ddot Scan(\ddot \sigma_{Points=32[sel=0.056]}(reports.Player))$
\item $\ddot Scan(\ddot \sigma_{Points=28[sel=0.096]}(reports.Player))$
\item $\ddot \chi_{list,name}(\ddot Scan(reports.Player))$
\item $\ddot \chi_{list,name}(\ddot Scan(\ddot \pi_{name, Points, Assists, Steals}(reports.Player)))$
\item $\ddot \chi_{list,name}(\ddot Scan(\ddot \sigma_{Points=28[sel=0.096]}(reports.Player)))$
\item $(team\_info \, \Join_{name} \, team\_to\_reports) \, \ddot \Join_{path, name} \, reports.Team$
\item $team\_stats \, \ddot \cup \, reports.Team$
\item $\ddot Scan(reports.Team)$
\item $(team\_info \, \Join_{name} \, team\_to\_reports) \, \ddot \Join_{path, name} \, \ddot \pi_{name, Wins, Losses, Total points}(reports.Team)$
\item $\pi_{name, Wins, Losses, Total points}(team\_stats) \, \ddot \cup \, \ddot \pi_{name, Wins, Losses, Total points}(reports.Team)$
\item $\ddot Scan(\ddot \pi_{name, Wins, Losses, Total points}(reports.Team))$
\item $(\sigma_{0.01}(team\_info) \, \Join_{name} \, team\_to\_reports) \, \ddot \Join_{path, name} \, reports.Team$
\item $(\sigma_{0.05}(team\_info) \, \Join_{name} \, team\_to\_reports) \, \ddot \Join_{path, name} \, reports.Team$
\item $(\sigma_{0.1}(team\_info) \, \Join_{name} \, team\_to\_reports) \, \ddot \Join_{path, name} \, reports.Team$
\item $\ddot Scan(\ddot \sigma_{Total points=99[sel=0.049]}(reports.Team))$
\item $\ddot Scan(\ddot \sigma_{Total points=102[sel=0.063]}(reports.Team))$
\item $\ddot \chi_{list,name}(\ddot Scan(reports.Team))$
\item $\ddot \chi_{list,name}(\ddot Scan(\ddot \pi_{name, Wins, Losses, Total points}(reports.Team)))$
\item $\ddot \chi_{list,name}(\ddot Scan(\ddot \sigma_{Total points=102[sel=0.063]}(reports.Team)))$
\end{enumerate}

\subsection{T-REx}

\begin{enumerate}
\item $nobel Personaltbl \, \ddot \cup \, nobel\_reports.Personal$
\item $nobel Careertbl \, \ddot \cup \, nobel\_reports.Career$
\item $nobel Careerinfo \, \ddot \Join_{path} \, nobel\_reports.Personal$
\item $nobel Personalinfo \, \ddot \Join_{path} \, nobel\_reports.Career$
\item $\ddot Scan(nobel\_reports.Personal)$
\item $\ddot Scan(nobel\_reports.Career)$
\item $nobel Personaltbl \, \ddot \cup \, \ddot \sigma_{country of citizenship=united states of america[sel=0.443]}(nobel\_reports.Personal)$
\item $nobel Careertbl \, \ddot \cup \, \ddot \sigma_{occupation=physicist[sel=0.127]}(nobel\_reports.Career)$
\item $\pi_{name, place of birth, country of citizenship}(nobel Personaltbl) \, \ddot \cup \, \ddot \pi_{name, place of birth, country of citizenship}(nobel\_reports.Personal)$
\item $\pi_{name, award received, educated at}(nobel Careertbl) \, \ddot \cup \, \ddot \pi_{name, award received, educated at}(nobel\_reports.Career)$
\item $\ddot \chi_{list,country of citizenship}(\ddot Scan(nobel\_reports.Personal))$
\item $\ddot \chi_{list,field of work}(\ddot Scan(nobel\_reports.Career))$
\end{enumerate}

\subsection{Aviation}

\begin{enumerate}
\item $(aircraft \, \Join_{aircraft\_registration\_number} \, aircraft\_to\_reports) \, \ddot \Join_{path, aircraft\_registration\_number} \, reports.incident$
\item $incidents \, \ddot \cup \, reports.incident$
\item $\ddot Scan(reports.incident)$
\item $(aircraft \, \Join_{aircraft\_registration\_number} \, aircraft\_to\_reports) \, \ddot \Join_{path, aircraft\_registration\_number} \, \ddot \pi_{location\_city, location\_state}(reports.incident)$
\item $\pi_{location\_city, location\_state}(incidents) \, \ddot \cup \, \ddot \pi_{location\_city, location\_state}(reports.incident)$
\item $\ddot Scan(\ddot \pi_{location\_city, location\_state}(reports.incident))$
\item $(\sigma_{0.3}(aircraft) \, \Join_{aircraft\_registration\_number} \, aircraft\_to\_reports) \, \ddot \Join_{path, aircraft\_registration\_number} \, reports.incident$
\item $(\sigma_{0.5}(aircraft) \, \Join_{aircraft\_registration\_number} \, aircraft\_to\_reports) \, \ddot \Join_{path, aircraft\_registration\_number} \, reports.incident$
\item $(\sigma_{0.8}(aircraft) \, \Join_{aircraft\_registration\_number} \, aircraft\_to\_reports) \, \ddot \Join_{path, aircraft\_registration\_number} \, reports.incident$
\item $incidents \, \ddot \cup \, \ddot \sigma_{location_state=colorado[sel=0.100]}(reports.incident)$
\item $incidents \, \ddot \cup \, \ddot \sigma_{weather_condition=visual meteorological conditions[sel=0.433]}(reports.incident)$
\item $incidents \, \ddot \cup \, \ddot \sigma_{aircraft_damage=destroyed[sel=0.667]}(reports.incident)$
\item $\ddot \chi_{list,aircraft\_damage}(\ddot Scan(reports.incident))$
\item $\ddot \chi_{list,location\_state}(\ddot Scan(reports.incident))$
\item $\ddot \chi_{list,weather\_condition}(\ddot Scan(reports.incident))$
\end{enumerate}

\subsection{Corona}

\begin{enumerate}
\item $corona\_stats \, \ddot \cup \, reports.summary$
\item $\pi_{new\_cases, new\_deaths, vaccinated}(corona\_stats) \, \ddot \cup \, \ddot \pi_{new\_cases, new\_deaths, vaccinated}(reports.summary)$
\item $\pi_{date, patients\_intensive\_care, twice\_vaccinated}(corona\_stats) \, \ddot \cup \, \ddot \pi_{date, patients\_intensive\_care, twice\_vaccinated}(reports.summary)$
\item $\pi_{date, incidence, vaccinated}(corona\_stats) \, \ddot \cup \, \ddot \pi_{date, incidence, vaccinated}(reports.summary)$
\item $\pi_{date, new\_cases, new\_deaths}(corona\_stats) \, \ddot \cup \, \ddot \pi_{date, new\_cases, new\_deaths}(reports.summary)$
\item $\pi_{new\_deaths, vaccinated, twice\_vaccinated}(corona\_stats) \, \ddot \cup \, \ddot \pi_{new\_deaths, vaccinated, twice\_vaccinated}(reports.summary)$
\item $\pi_{date, vaccinated, twice\_vaccinated}(corona\_stats) \, \ddot \cup \, \ddot \pi_{date, vaccinated, twice\_vaccinated}(reports.summary)$
\item $\pi_{new\_deaths, incidence, vaccinated}(corona\_stats) \, \ddot \cup \, \ddot \pi_{new\_deaths, incidence, vaccinated}(reports.summary)$
\item $\pi_{new\_cases, new\_deaths, incidence}(corona\_stats) \, \ddot \cup \, \ddot \pi_{new\_cases, new\_deaths, incidence}(reports.summary)$
\item $\pi_{date, new\_cases, patients\_intensive\_care, vaccinated}(corona\_stats) \, \ddot \cup \, \ddot \pi_{date, new\_cases, patients\_intensive\_care, vaccinated}(reports.summary)$
\item $\pi_{new\_cases, new\_deaths, incidence, twice\_vaccinated}(corona\_stats) \, \ddot \cup \, \ddot \pi_{new\_cases, new\_deaths, incidence, twice\_vaccinated}(reports.summary)$
\item $\pi_{date, new\_cases, new\_deaths, patients\_intensive\_care}(corona\_stats) \, \ddot \cup \, \ddot \pi_{date, new\_cases, new\_deaths, patients\_intensive\_care}(reports.summary)$
\item $\pi_{date, new\_cases, vaccinated, twice\_vaccinated}(corona\_stats) \, \ddot \cup \, \ddot \pi_{date, new\_cases, vaccinated, twice\_vaccinated}(reports.summary)$
\item $\pi_{date, new\_deaths, patients\_intensive\_care, twice\_vaccinated}(corona\_stats) \, \ddot \cup \, \ddot \pi_{date, new\_deaths, patients\_intensive\_care, twice\_vaccinated}(reports.summary)$
\item $\pi_{date, new\_deaths, incidence, vaccinated}(corona\_stats) \, \ddot \cup \, \ddot \pi_{date, new\_deaths, incidence, vaccinated}(reports.summary)$
\end{enumerate}

\end{document}